\RequirePackage{ifpdf}
\documentclass[11pt,a4paper]{article}
\usepackage{epsfig,jheppubme}

\usepackage{multirow,longtable,enumerate}

\usepackage{physics}
\usepackage{tensor}
\usepackage{amsmath}
\usepackage{amssymb}
\usepackage{appendix}
\usepackage{mathtools}
\usepackage{mathrsfs}
\usepackage[normalem]{ulem}
\usepackage{tikz}
\usepackage{booktabs}
\usepackage{bbm}
\usepackage{verbatim}

\font\cmss=cmss12

\newcommand{\ovl}{\overline}

\newcommand\half{\frac12}

\newcommand\bi{\begin{itemize}}
\newcommand\ei{\end{itemize}}

\newcommand\tq{{\tilde q}}

\newcommand\tc{{\tilde c}}

\newcommand\ta{{\tilde a}}
\newcommand\tb{{\tilde b}}

\newcommand\tp{{\tilde p}}

\newcommand\tilh{{\tilde h}}
\newcommand\tchi{{\widetilde \chi}}

\newcommand{\tepsilon}{{\tilde\epsilon}}
\newcommand{\tmu}{{\tilde\mu}}

\newcommand\bea{\begin{eqnarray}}
\newcommand\eea{\end{eqnarray}}
\newcommand\be{\begin{equation}}
\newcommand\ee{\end{equation}}

\newcommand\bz{{\bar z}}
\newcommand\ba{{\bar a}}
\newcommand\bb{{\bar b}}
\newcommand\bg{{\bar g}}
\newcommand\bta{{\bar {\tilde a}}}
\newcommand\btb{{\bar {\tilde b}}}

\newcommand\btq{{\bar {\tilde q}}}
\newcommand\btau{{\bar \tau}}

\newcommand\cG{{\cal G}}
\newcommand\cR{{\cal R}}

\newcommand\cF{{\cal F}}

\newcommand\cS{{\cal S}}

\newcommand\m{{\mu}}

\newcommand\tcF{{\widetilde{\cal F}}}

\newcommand\tD{{\widetilde D}}

\newcommand\sfrac[2]{{\textstyle\frac{#1}{#2}}}

\newcommand\shalf{{\textstyle\frac12}}

\newcommand\ZZ{\hbox{Z\kern-.4emZ}}
\newcommand\sZZ{\hbox{\sevenfont Z\kern-.4emZ}}

\newcommand{\eref}[1]{Eq.\,(\ref{#1})}
\newcommand{\Comment}[1]{{}}

\newcommand{\wtd}{\widetilde}

\newcommand{\mc}{\mathcal}

\newcommand{\mfs}{\mathfrak s}

\allowdisplaybreaks

\def\IB{\relax{\rm I\kern-.18em B}}
\def\IC{{\relax\hbox{\kern.3em{\cmss I}$\kern-.4em{\rm C}$}}}
\def\ID{\relax{\rm I\kern-.18em D}}
\def\IE{\relax{\rm I\kern-.18em E}}
\def\IF{\relax{\rm I\kern-.18em F}}
\def\II{\relax{\rm I\kern-.18em I}}

\def\Id{\relax{1\kern-.32em 1}}
\def\IG{\relax\hbox{$\inbar\kern-.3em{\rm G}$}}
\def\IR{\relax{\rm I\kern-.18em R}}

\newcommand{\eps}{\epsilon}

\title{Universal Correlators and Novel Cosets in 2d RCFT} 

\author[a]{Sunil Mukhi\,}

\author[b]{and Rahul Poddar\,}

\affiliation[a]{Indian Institute of Science Education and Research,\\ \it Homi Bhabha Rd, Pashan, Pune 411 008, India}
\affiliation[b]{Science Institute, University of Iceland,\\
 Dunhaga 3, Reykjav\'ik 107, Iceland} 

\emailAdd{sunil.mukhi@gmail.com}
\emailAdd{rahul.poddar.305@gmail.com}

\abstract{The two-character level-1 WZW models corresponding to Lie algebras in the Cvitanovi\'c-Deligne series $A_1,A_2,G_2,D_4,F_4,E_6,E_7$ have been argued to form coset pairs with respect to the meromorphic $E_{8,1}$ CFT. Evidence for this has taken the form of holomorphic bilinear relations between the  characters. We propose that suitable 4-point functions of primaries in these models also obey bilinear relations that combine them into current correlators for $E_{8,1}$, and provide strong evidence that these relations hold in each case. Different cases work out due to special identities involving tensor invariants of the algebra or hypergeometric functions. In particular these results verify previous calculations of correlators for exceptional WZW models, which have rather subtle features. We also find evidence that the intermediate vertex operator algebras A$_{0.5}$ and E$_{7.5}$, as well as the three-character A$_{4,1}$ theory, also appear to satisfy the novel coset relation.}

\preprint{}


\arxivnumber{}

\begin{document}

\maketitle

\section{Introduction}
\label{Intro}

The finite-dimensional Lie algebras $A_{1}$, $A_{2}$,  $G_{2}$, $D_{4}$, $F_{4}$, $E_{6}$, $E_{7}$, $E_{8}$ were shown, many years ago by Cvitanovi\'c and by Deligne \cite{Cvitanovic:1977ym,  Deligne:1, Deligne:2, Cvitanovic:2008zz}, to belong in a special series with remarkable mathematical properties. These include miraculous formulae for the dimensions of representations, parametrised in terms of the dual Coxeter number. Later it was argued  by Landsberg and Manivel \cite{Landsberg:2004} that there is a ``hole'' in this series between $E_7$ and $E_8$ which suggests the existence of a new structure (not a normal Lie algebra) denoted $E_{7.5}$. We will refer to the full collection as the CDLM series of Lie Algebras.

On the physics side, the classification of Rational Conformal Field Theories (RCFT) by their number of conformal blocks was initiated by Mathur et al in \cite{Mathur:1988na,Mathur:1988gt} using a bootstrap procedure to implement modularity of the characters together with positive integrality of their coefficients. All two-character CFT's with a vanishing Wronskian index were classified in these papers, making up what is now known as the MMS series. These consist of the WZW models $A_{1,1}$, $A_{2,1}$, $G_{2,1}$, $D_{4,1}$, $F_{4,1}$, $E_{6,1}$, $E_{7,1}$, $E_{8,1}$ where the subscripts label the rank and level, as well as two outliers which formally give rise to negative fusion rules as computed by the Verlinde formula \cite{Verlinde:1988sn}. One of these outliers was identified in \cite{Mathur:1988na}, after interchanging characters, with the famous non-unitary Lee-Yang minimal CFT, which provides a nice physical realisation. However from the mathematical point of view it is better to keep the ``standard'' order of characters (in which the central charge parameter is positive) and it was subsequently proposed by Kawasetsu \cite{Kawasetsu:2014} that in this form, both the outliers can be identified as Intermediate Vertex Operator Algebras (IVOA). These  generalise the usual axioms of RCFT, and the two outliers above are now known as the IVOA's $A_{0.5}$ and $E_{7.5}$. 

Thus, the CDLM series of (generalised) Lie algebras, obtained using miraculous features of representation theory, also emerges -- from a completely distinct starting point -- as the MMS series of (generalised) RCFT obtained via the modular bootstrap\footnote{Except for A$_{0.5}$ which presumably corresponds to a trivial Lie algebra.}. The special mathematical properties of the CDLM series, elucidated in the mathematical literature, seem to have a counterpart in the special RCFT properties of having two characters and vanishing Wronskian index\footnote{As far as we know, this correspondence has never been explained.}. 

An intriguing feature of the MMS series is that the various theories pair up such that the central charges $(c,\tc)$ of a pair add up to 8 and the conformal dimensions of the non-trivial primaries $(h,\tilh)$ add up to 1. The pairs are as follows:
\be
\begin{tabular}{lll}
$(A_{1,1},E_{7,1})$ : & $(c,\tc)=(1,7)$, & $(h,\tilh)=(\sfrac14,\sfrac34)$\\
$(A_{2,1},E_{6,1})$ : & $(c,\tc)=(2,6)$, & $(h,\tilh)=(\sfrac13,\sfrac23)$\\
$(G_{2,1},F_{4,1})$ : & $(c,\tc)=(\sfrac{14}{5},\sfrac{26}{5})$, & $(h,\tilh)=(\sfrac25,\sfrac35)$ \\
$(D_{4,1},D_{4,1})$ : & $(c,\tc)=(4,4)$, & $(h,\tilh)=(\sfrac12,\sfrac12)$\\
$(A_{0.5},E_{7.5})$ : & $(c,\tc)=(\sfrac{2}{5},\sfrac{38}{5})$, & $(h,\tilh)=(\sfrac15,\sfrac45)$
\end{tabular}
\label{chpairing}
\ee
It is noteworthy that this relation covers both simply-laced and non-simply-laced cases as well as the IVOA case. Also one sees that $D_{4,1}$ is paired up with itself.

An explanation for these numerical facts was found in \cite{Gaberdiel:2016zke} which proposed a novel coset construction for RCFT's. While the usual coset procedure \cite{Goddard:1984vk} starts with multi-character WZW models and takes the coset of one by another to obtain non-WZW models (such as Virasoro minimal models), the novel coset construction of \cite{Gaberdiel:2016zke}  starts with a {\em meromorphic} CFT (typically not a WZW model) and takes its coset by a WZW model to get new and interesting CFT's with (in general) a larger Wronskian index.  The coset relation is embodied in a holomorphic bilinear pairing between the characters of the denominator theory and the coset, combining them into the single modular-invariant character of the numerator meromorphic theory. This novel coset procedure has been used in recent times to construct previously unknown families of two-character RCFT with arbitrarily large Wronskian index \cite{Chandra:2018ezv}.

While the main examples in \cite{Gaberdiel:2016zke} dealt with meromorphic CFT with $c=24$ (several interesting examples with $c=32$ were found more recently \cite{Chandra:2018ezv}), it was noted that the same construction also pairs the MMS series characters into the character of the meromorphic CFT $E_{8,1}$ \footnote{While this theory formally emerges in the MMS analysis of two-character CFT, it is of course well-known to be a meromorphic (single-character) CFT whose character is modular invariant up to a phase.}. Thus, if $(\chi_i(\tau),\tchi_i(\tau))$ are the characters of a pair of MMS-series theories as in \eref{chpairing}, they satisfy the holomorphic relation:
\be
\sum_{i=0}^1\chi_i(\tau)\tchi_i(\tau)=j(\tau)^\frac13=\chi_{E_{8,1}}(\tau)
\label{charpairing}
\ee
where $j(\tau)$ is the Klein modular invariant. Among other things, the above relation equates the modular $\cS$-matrix of the two theories in a coset pair.  

Given that such a pairing exists between characters, one may ask if a similar pairing holds for four-point functions. This is the subject of the present work. We will consider the four-point correlators of the non-trivial primary for each member of the MMS series and argue that when the holomorphic conformal blocks are multiplied with each other, they satisfy a bilinear relation analogous to \eref{charpairing} where the RHS is the four-point current correlator of $E_{8,1}$. This relation will turn out to be considerably more involved than the bilinear pairing of characters. The reason is that it intimately involves structure constants and invariant tensors of Lie algebras and the fusion rules of the corresponding theories. By contrast, the characters merely count states and \eref{charpairing} (despite being of course non-trivial) only relates the state-counting of a pair of theories to that of $E_{8,1}$ without specifically invoking group theory. 

The 4-point correlators for WZW models based on classical Lie algebras are quite well-known, starting with the pioneering work of \cite{Knizhnik:1984nr} based on the null vectors of the RCFT. This method does not, however, seem to have been applied to exceptional WZW models\footnote{Here we mean WZW models whose Kac-Moody algebra is based on an exceptional Lie algebra, as opposed to exceptional invariants for A$_1$ WZW models -- which seem to be far more widely studied.}. Instead, the problem has been tackled using the Wronskian method for correlators of RCFT originally proposed in \cite{Mathur:1988rx,Mathur:1988gt} -- specifically, the correlator of $F_{4,1}$ WZW theory was calculated in \cite{Mathur:1988gt} and the Knizhnik-Zamolodchikov equation was derived for the fundamental primaries of $G_2$ and $F_4$ in \cite{Fuchs:1989pp}. More recently, a master formula was obtained in \cite{Mukhi:2017ugw} for the 4-point function of primaries for all members of the MMS series described above\footnote{The IVOA's $A_{0.5}$ and $E_{7.5}$ were not specifically discussed there, though we will find that the master formula applies to them as well.}. 

Four-point correlation functions in RCFT are expressed as a sum over the modulus-squared of  conformal blocks that are holomorphic away from coincident points. Each block corresponds to the flow of a different conformal family between pairs of fields, and the blocks have monodromies under exchange of these fields. While this much is true for minimal models and widely known, correlators for WZW models have an additional non-trivial feature. Each primary has a degeneracy equal to the dimension of the representation of the correponding finite Lie algebra. Hence the conformal blocks themselves are decomposed over all possible independent tensor invariants that can contribute, and there is a sub-block for each invariant. Some unusual selection rules are imposed by group theory over and above those given by CFT, which tell us that in some situations the primary itself is not the leading contribution to a given conformal block, but instead the leading contribution comes from some definite secondary over that primary. This was noted in \cite{Mathur:1988rx,Mathur:1988gt} (the former reference considers A$_{n,1}$ correlators while the latter describes the four-point function of $F_{4,1}$ as an example) and discussed in some detail in \cite{Fuchs:1989pp,Fuchs:1992te}.

The more detailed and comprehensive results of \cite{Mukhi:2017ugw} encountered the fact -- well-known in representation theory, and noted previously in \cite{Fuchs:1989pp} -- that for exceptional algebras, the tensor product of the fundamental representation with itself does not simply give irreducible symmetric and antisymmetric parts. Instead each of these breaks up further into irreps. As a result a 4-point correlator in the exceptional case may have to be decomposed into as many as five or six different tensor structures (not just two, as for A$_{n,1}$) and the sub-blocks calculated separately in each case. While the tensor structures were not explicitly identified in \cite{Mathur:1988gt, Mukhi:2017ugw}, they were labelled and rules were obtained to compute the contribution for each of them. Meanwhile \cite{Fuchs:1989pp} focused on deriving the differential equation and did not explicitly write the solutions, which are more relevant to the present discussion. 

Hence, here we will first fill in a few details that were missing in the previous works. In particular we will discuss the tensor invariants appearing in the correlations functions of the MMS series in considerable detail. 
We then proceed to consider bilinear products of conformal blocks. We will find that the calculations, which are highly sensitive to the above subtleties, hold up very well and the existence of a bilinear relation among correlators of coset-paired theories is supported by strong evidence in each case. In no case do we find any term that would contradict it. We will also find that in some cases, non-trivial relations between them are crucial to enable the proposed bilinear coset relation. Thereby we (i) provide a precise statement of what the novel coset construction implies for correlators, (ii) find evidence for this statement,  This can also be taken as supporting evidence for the correctness of the formulae in \cite{Mukhi:2017ugw} where a couple of slightly ad-hoc assumptions had to be made. 

We go beyond the MMS series to consider the three-character pair $(A_{4,1},A_{4,1})$ which also forms a coset pair and satisfies the bilinear relation \eref{charpairing}. Here too we find convincing evidence for a bilinear relation between conformal blocks. Like the pair $(D_{4,1},D_{4,1})$, this is a relatively simple case, since exceptional algebras are not involved.

Let us mention here that for all simply-laced pairs, the conformal blocks are algebraic functions of the cross-ratio, and the existence of a bilinear relation between blocks, while highly non-trivial, is perhaps not too miraculous. However for the pair $(G_{2,1},F_{4,1})$ the blocks are hypergeometric functions that cannot be simplified into algebraic functions. But when we multiply them to find a bilinear relation then the result remarkably simplifies into a purely algebraic one due to identities among hypergeometric functions. Such a simplification is of course essential for the bilinear relation to hold, given that the current correlator of E$_{8,1}$ appears on the right-hand-side. An added bonus is that, with some minimal assumptions,  a similar result holds for the pair of IVOA's $(A_{0.5},E_{7.5})$. This appears to give us some novel information about IVOA's, whose correlators have not been previously considered as far as we know.

The plan of this paper is as follows. In Section \ref{WZWrev} we review relevant results about primary correlators and current correlators in WZW models, and state the novel coset conjecture as it applies to 4-point functions. In Section \ref{testingsec} discuss each coset pair in the MMS series, providing evidence for the bilinear relation in each case. At the end we briefly discuss the case of A$_{4,1}\oplus$ A$_{4,1}$ which is not part of the MMS series (these are three-character theories) but also satisfies a similar coset relation. In Section \ref{conclusionsec} we conclude with a summary of what has been achieved and what remains to be done. In Appendix \ref{App.Bilinear} we state and prove a bilinear identity between hypergeometric functions,  and in Appendix \ref{App.Bilinear.Char} we use it to prove the bilinear relation among characters.

\section{Review of WZW Correlators and the Novel Coset Conjecture}
\label{WZWrev}

Let us summarise the ingredients in the computation of four-point functions for WZW models. We begin by describing primary correlators. Subsequently we write down four-point correlators of Kac-Moody currents and state the novel coset conjecture for correlation functions.

\subsection{Primary Correlators}

\label{primcorr}

Let us pick a WZW model based on a simple Kac-Moody algebra and consider primaries $g_{ab}(z,\bz)$ in some representation $A$, having conformal dimensions $h_A,\bar h_A$. Assuming the representation to be complex, we consider the correlator:
\be
\cG_{a_1\ba_2\ba_3a_4;b_1\bb_2\bb_3b_4}(z_i,\bz_i)=
\langle g_{a_1b_1}(z_1,\bz_1)\,\bg_{\ba_2\bb_2}\,(z_2,\bz_2)\,\bg_{\ba_3\bb_3}(z_3,\bz_3)\,g_{a_4b_4}(z_4,\bz_4)\rangle
\label{fullcorr.1}
\ee
where the barred entries correspond to the complex-conjugate representation. The correlator is conveniently expressed in terms of the cross-ratio, for which we first write:
\be
\cG_{a_1\ba_2\ba_3a_4;b_1\bb_2\bb_3b_4}(z_i,\bz_i)=(z_{14}z_{32}\bz_{14}\bz_{32})^{-2h_A}G_{a_1\ba_2\ba_3a_4;b_1\bb_2\bb_3b_4}(z,\bz)
\label{corr.cross}
\ee
where $z=\frac{z_{12}z_{34}}{z_{14}z_{32}}$, and 
then take the limits $z_2,z_3,z_4\to0,1,\infty$. 

Next the correlator $G$ is expressed as a sum over conformal blocks:
\be
G_{a_1\ba_2\ba_3a_4;b_1\bb_2\bb_3b_4}(z,\bz)=\sum_\alpha
\cF_{\alpha,a_1\ba_2\ba_3a_4}(z)\bar \cF_{\alpha,b_1\bb_2\bb_3b_4}(\bz)
\ee
In the present work we will mostly restrict our attention to the case where $\alpha\in \{1,2\}$, which is the case for all MMS-series theories -- though in some of them the $\alpha=2$ term will decouple and we then have only one conformal block. The extra case of A$_{4,1}$ has in principle three blocks, but again all but the first one decouple -- this is a general feature of the A$_{n,1}$ theories.

As we see, the holomorphic blocks carry indices corresponding to the representations of the fields in the correlator. The correlator is nonzero only when these indices combine into a singlet\footnote{This is purely due to the finite-dimensional zero-mode Lie algebra that is a subalgebra of the Kac-Moody algebra.}. Hence the dependence of the blocks on these indices must be through invariant tensors of the algebra corresponding to all possible ways of combining the four representations into a singlet. Thus the blocks can, in turn, be written in terms of a set of sub-blocks with no Lie algebra indices, each multiplied by an independent invariant tensor:
\be
\cF_{\alpha,a_1\ba_2\ba_3a_4}(z)=\sum_{R_p\in R_A\otimes R_A} D^{(p)}_{a_1\ba_2\ba_3a_4}\cF_\alpha^{(p)}(z)
\label{subblocks}
\ee
Four-point conformal blocks $\cF_\alpha$ are labelled by the conformal families flowing in any chosen channel: this can be $(a_1\ba_2), (a_1\ba_3)$ or $(a_1a_4)$. However, the $(a_1a_4)$ channel is most convenient because both indices are in the same representation and so the invariant tensors are either symmetric or antisymmetric. 
This was the choice made in \cite{Mathur:1988rx, Mathur:1988gt}. Thus, the conformal families that label the blocks correspond to the possible outputs of $\cR_A\otimes \cR_A$. However, at low values of the Kac-Moody level there will be a truncation and we will only see the integrable representations.

Due to our choice of channel, these blocks are invariant (up to an overall phase) under the interchange $z\to 1-z$, $a_1\to a_4$. Thus, the sub-block $\cF_\alpha^{(p)}$ should be symmetric/antisymmetric under $z\to 1-z$ if the corresponding invariant tensor $D^{(p)}_{\alpha\,a_1\ba_2\ba_3a_4}$ that it multiplies is symmetric/antisymmetric under $a_1\leftrightarrow a_4$. This leads us to consider symmetric and antisymmetric $D^{(p)}$ separately:
\be
\begin{split}
D^{\rm S}&=\shalf(\delta_{a_1\ba_2}\delta_{\ba_3a_4}+\delta_{a_1\ba_3}\delta_{\ba_2a_4})\\
D^{\rm A}&=\shalf(\delta_{a_1\ba_2}\delta_{\ba_3a_4}-\delta_{a_1\ba_3}\delta_{\ba_2a_4}) 
\end{split}
\label{DSDA}
\ee
Each of these corresponds to a sum of certain representations flowing in the $a_1a_4$ channel. Depending on the relevant Lie algebra, the above invariants may themselves correspond to irreducible representations or may be further reducible, as we will see case by case. Whenever these are irreducible representations, the above tensors are precisely the $D^{(p)}$ of \eref{subblocks} with $p$ ranging over 0 and 1, with 0 by convention being the symmetric channel and 1 the antisymmetric. In more general cases, each of $D^{\rm S}$ and $D^{\rm A}$ will break up into sums over a set of $D^{(p)}$. 

An important point, first highlighted in \cite{Mathur:1988rx} and investigated further in \cite{Mathur:1988gt,Fuchs:1989pp,Fuchs:1992te}, is that generically there is an interplay between the index $\alpha$ labelling the conformal family in the $a_1a_4$ channel, and the index $p$ labelling the representation of the leading (lowest-dimension) state that actually flows in that channel. As a concrete example, if we consider the block $\cF_\alpha$ where $\alpha$ labels a primary arising in the symmetric part of the product $\cR_A\otimes\cR_A$, then the primary $\alpha$ itself cannot flow if we are in the sub-block corresponding to an antisymmetric tensor structure $D^{\rm A}$. Instead, the lowest-dimension state that flows will be some particular secondary. This point will be relevant in what follows.

Another point to note is that sometimes the representation of $\phi$ will be real or pseudo-real. In the real cases we will of course consider the four-point function of the same (real) field. In the pseudo-real case, the field and its complex conjugate are equal only after a linear transformation. Here we will find it notationally simpler to let all fields in the correlator be the same, rather than the middle two being complex conjugates. 
However then contraction of a representation with itself has to be done via an antisymmetric tensor (as is familiar for A$_1$). In all these cases, the correlator \eref{fullcorr.1} reduces to:
\be
\cG_{a_1a_2a_3a_4;b_1b_2b_3b_4}(z_i,\bz_i)=
\langle g_{a_1b_1}(z_1,\bz_1)\,g_{a_2b_2}\,(z_2,\bz_2)\,g_{a_3b_3}(z_3,\bz_3)\,g_{a_4b_4}(z_4,\bz_4)\rangle
\label{fullcorr.real}
\ee
This is crossing-symmetric under, for example, the exchange of 1 with 4:
\be
\cG_{a_4a_2a_3a_1;b_4b_2b_3b_1}(z_4,z_2,z_3,z_1;\bz_4,\bz_2,\bz_3,\bz_1)
=
\cG_{a_1a_2a_3a_4;b_1b_2b_3b_4}(z_1,z_2,z_3,z_4;\bz_1,\bz_2,\bz_3,\bz_4)
\ee
and similarly for all other pairwise exchanges.

In the complex case things are slightly different. We have crossing symmetry under the exchange of 1 with 4 (both holomorphic) in \eref{fullcorr.1}:
\be
\cG_{a_4\ba_2\ba_3a_1;b_4\bb_2\bb_3b_1}(z_4,z_2,z_3,z_1;\bz_4,\bz_2,\bz_3,\bz_1)
=
\cG_{a_1\ba_2\ba_3a_4;b_1\bb_2\bb_3b_4}(z_1,z_2,z_3,z_4;\bz_1,\bz_2,\bz_3,\bz_4)
\ee
and similarly for 2 with 3 (both anti-holomorphic). Other exchanges give rise to correlators with different placements of holomorphic and anti-holomorphic indices. Fortunately, when we verify the coset relation for such cases, we will automatically find a sum over all placements of the indices and this will turn out to give the expected answers.

\subsection{Current Correlators and the Conjecture}

The current-current OPE of a WZW model is:
\begin{equation}
  J_A (z) J_B (w) = \frac{k\delta_{AB}}{(z-w)^2} + \frac{i f_{ABC} J_C(w)}{(z-w)}
\end{equation}
To find current correlators, we can use the OPE to construct a recursion relation
\begin{equation}
  \begin{split}
  &\ev{J_{A_1}(z_1) \cdots J_{A_n}(z_n) } \\
  &\quad= \sum_{j = 2}^{n} \ev{J_{A_2}(z_2)\cdots J_{A_{j-1}}(z_{j-1}) \left(\frac{k\delta_{A_1A_j}}{(z_1-z_j)^2} + \frac{i f_{A_1A_jB} J_B(z_j)}{(z_1-z_j)}\right)J_{A_{j+1}}(z_{j+1})\cdots J_{A_{n}}(z_{n})}
  \end{split}
\end{equation}
Using this, we have:
\begin{align}
  \ev{J_{A_1}(z)J_{A_2}(w)} &= \frac{k\delta_{A_1A_2}}{(z-w)^2}\\
  \ev{J_{A_1}(z_1)J_{A_2}(z_2)J_{A_3}(z_3)} &= \frac{ikf_{A_1A_2A_3}}{z_{12}z_{13}z_{23}}\\
  \begin{split}
  \ev{J_{A_1}(z_1)J_{A_2}(z_2)J_{A_3}(z_3)J_{A_4}(z_4)} &=
  \frac{k^2\delta_{A_1A_2}\delta_{A_3A_4}}{(z_{12})^2 (z_{34})^2}
  + \frac{k^2\delta_{A_1A_3}\delta_{A_2A_4}}{(z_{13})^2 (z_{24})^2}
  + \frac{k^2\delta_{A_2A_3}\delta_{A_1A_4}}{(z_{14})^2(z_{23})^2 } \\ 
  &\;- k\left(
    \frac{f_{A_1A_2B}f_{BA_3A_4}}{z_{12}z_{23}z_{24}z_{34}}
  + \frac{f_{A_1A_3B}f_{BA_4A_2}}{z_{13}z_{23}z_{24}z_{34}}
  + \frac{f_{A_1A_4B}f_{BA_2A_3}}{z_{14}z_{23}z_{24}z_{34}}
  \right)
  \end{split}
\end{align}
where $z_{ij} = z_i - z_j$. 
The structure constants obey the Jacobi identity:
\begin{equation}
  f_{A_1A_2B}f_{BA_3A_4}+ f_{A_1A_3B}f_{BA_4A_2}+  f_{A_1A_4B}f_{BA_2A_3} = 0
\end{equation}
These correlators are symmetric under the interchange $A_i\leftrightarrow A_j, z_i\leftrightarrow z_j$. In the above expression for the 4-point function, symmetry under pairwise interchanges between any of $(A_2,z_2)$, $(A_3,z_3)$, $(A_4,z_4)$ is manifest, but to see the symmetry under exchange of $(A_1,z_1)$ with any of these one has to use the Jacobi identity.

\Comment{
The second line of the 4 point function is not manifestly symmetric.
However, using the Jacobi identity, we can show that it is symmetric on exchange of any two currents in the four point function.
First we pull out the common factors and then we use the Jacobi identity

\begin{equation}
  \begin{split}
    &\quad\frac{1}{z_{23}z_{24}z_{34}} \left( \frac{f^{abe}f^{ecd}}{z_{12}} + \frac{f^{ace}f^{edb}}{z_{13}} + \frac{f^{ade}f^{ebc}}{z_{14}} \right) \\
    &= \frac{1}{z_{23}z_{34}z_{24}} \left( f^{abe}f^{ecd}\left(\frac{1}{z_{12}}-\frac{1}{z_{14}}\right) + f^{ace}f^{edb}\left(\frac{1}{z_{13}} - \frac{1}{z_{14}}\right) \right) \\
    &= \frac{1}{z_{23}z_{34}z_{24}} \left( f^{abe}f^{ecd}\left(\frac{z_{24}}{z_{12}z_{14}}\right) + f^{ace}f^{edb}\left(\frac{z_{34}}{z_{13}z_{14}}\right) \right) \\
    &= \frac{1}{z_{23}z_{14}} \left( f^{abe}f^{ecd}\left(\frac{1}{z_{12}z_{34}}\right) + f^{ace}f^{edb}\left(\frac{1}{z_{13}z_{24}}\right) \right) \\
    &= \frac{1}{z_{14}}\left( \frac{ f^{abe}f^{ecd}}{z_{34}} \left(\frac{1}{z_{12}z_{23}}\right) + \frac{f^{ace}f^{edb}}{z_{13}}\left(\frac{1}{z_{24}z_{23}}\right) \right) \\
    &= \frac{1}{z_{14}}\left( \frac{ f^{abe}f^{ecd}}{z_{34}z_{13}} \left(\frac{1}{z_{12}}+\frac{1}{z_{23}}\right) + \frac{f^{ace}f^{edb}}{z_{13}z_{43}}\left(\frac{1}{z_{24}}-\frac{1}{z_{23}}\right) \right) \\
    &= -\frac{1}{z_{14}z_{13}z_{34}}\left(\frac{f^{abe}f^{ecd}}{z_{21}} + \frac{f^{ace}f^{edb}}{z_{24}} + \frac{f^{ade}f^{ebc}}{z_{23}} \right) \\
    &= \frac{1}{z_{14}z_{13}z_{34}}\left(\frac{f^{bae}f^{ecd}}{z_{21}} + \frac{f^{bce}f^{ade}}{z_{23}} + \frac{f^{bde}f^{eac}}{z_{24}}  \right) \\
  \end{split}
\end{equation}
The last line is also obtained by switching $z_1, z_2$ and $a,b$ simultaneously.
Since the ``original'' term is symmetric in pairwise exchange of $(z_2,b), (z_3,c), (z_4,d)$, this suffices to show that the term is completely symmetric under exchange of all 4 pairs $(z_i, a_i)$.
}

We now set the points $z_1 = z, z_2 = 0, z_3 = 1$ and use a conformal transformation to set $z_4 = \infty$. We will use the current correlator exclusively for E$_{8,1}$ so at this stage we specialise to that case, and set the level $k=1$. Thus the E$_{8,1}$ current correlator is:
\begin{equation}
  \begin{split}
    \ev{J_{A_1}J_{A_2} J_{A_3} J_{A_4} }(z) 
    &=  \frac{\delta_{A_1A_2}\delta_{A_3A_4}}{z^2}
  + \frac{\delta_{A_1A_3}\delta_{A_2A_4}}{(1-z)^2}
  + \delta_{A_1A_4}\delta_{A_2A_3} \\
& \quad +  \frac{f_{A_1A_2B}f_{BA_3A_4}}{z}+\frac{f_{A_1A_3B} f_{BA_2A_4}}{1-z}
  \end{split}
\label{currentcorr}
\end{equation}
It will be useful for what follows to highlight the symmetry properties of this correlator. It is symmetric under the simultaneous interchange $A_2\leftrightarrow A_3$ and $z\leftrightarrow 1-z$. Also it is symmetric under exchange of the pair $A_2, A_3$ with the pair $A_1,A_4$. Next, focusing on the residues of pole terms as $z\to 0$, we see that the residue of the double pole is symmetric under $A_3\leftrightarrow A_4$ and that of the single pole is antisymmetric under the same exchange. Finally, the latter satisfies a Jacobi identity: the residue of the single pole, when summed over cyclic permutations of $A_2,A_3,A_4$, vanishes.

The Kac-Moody algebra $E_{8,1}$ can be decomposed into subalgebras corresponding to direct sums of pairs A$_{1,1}\oplus$ E$_{7,1}$, A$_{2,1}\oplus$ E$_{6,1}$, G$_{2,1}\oplus$ F$_{4,1}$, D$_{4,1}\oplus$ D$_{4,1}$, A$_{4,1}\oplus$ A$_{4,1}$. All of these are maximal subalgebras except D$_{4,1}\oplus$ D$_{4,1}$ which is contained in D$_{8,1}$ which itself is a subalgebra of E$_{8,1}$. Under the above decompositions, the adjoint of $E_{8,1}$, of dimension 248, breaks up as: 
\be
\begin{split}
{\rm A}_{1,1}\oplus {\rm E}_{7,1}\!: ~\mathbf{248} &\to (\mathbf{3},\mathbf{1})+(\mathbf{2},\mathbf{56})+(\mathbf{1},\mathbf{133})\\
{\rm A}_{2,1}\oplus {\rm E}_{6,1}\!: ~\mathbf{248} &\to (\mathbf{8},\mathbf{1})+(\mathbf{3},\mathbf{27})+(\mathbf{\overline{3}},\mathbf{\overline{27}})+(\mathbf{1},\mathbf{78})\\
{\rm G}_{2,1}\oplus {\rm F}_{4,1}\!: ~\mathbf{248} &\to (\mathbf{14},\mathbf{1})+(\mathbf{7},\mathbf{26})+(\mathbf{1},\mathbf{52})\\
{\rm D}_{4,1}\oplus {\rm D}_{4,1}\!: ~\mathbf{248} &\to (\mathbf{28},\mathbf{1})+(\mathbf{8_v},\mathbf{8_v})+(\mathbf{8_s},\mathbf{8_c})+(\mathbf{8_c},\mathbf{8_s})
+(\mathbf{1},\mathbf{28})\\
{\rm A}_{4,1}\oplus {\rm A}_{4,1}\!: ~\mathbf{248} &\to (\mathbf{24},\mathbf{1})+(\mathbf{5},\mathbf{10})+(\mathbf{\overline{5}},\mathbf{\overline{10}})+(\mathbf{5},\mathbf{\overline{10}})+(\mathbf{\overline{5}},\mathbf{10})+ (\mathbf{1},\mathbf{24})
\end{split}
\label{decomp}
\ee
As mentioned in the Introduction, the characters of the above pairs obey a holomorphic bilinear relation to the characters of E$_{8,1}$. This was stated in \cite{Gaberdiel:2016zke} but a proof was not provided there. Hence we provide this here, after expressing the characters in terms of hypergeometric functions following \cite{Mathur:1988gt,Naculich:1988xv}. This is done in Appendix \ref{App.Bilinear.Char}, using results on hypergeometric functions derived in Appendix \ref{App.Bilinear}.

We now conjecture an analogous relation between 4-point current correlators of $E_{8,1}$ on the plane and products of conformal blocks for 4-point functions of fundamental primaries for the above pairs. Let the conformal blocks of the first member of a pair be denoted $\cF_{\alpha,a_1\ba_2\ba_3a_4}(z)$ and those of the second member be $\tcF_{\alpha,\ta_1\bta_2\bta_3\ta_4}(z)$. The $E_{8,1}$ currents  are labelled $J_{A_i}$ as above. We now restrict all the $A_i$ to lie in the set corresponding to  one of the middle terms on the RHS of \eref{decomp}, which involve a fundamental representation of each member of the pair. These restricted indices can be thought of as composite: $A_i=(a_i\ta_i)$ for $i=1,2,3,4$ where $a_i,\ta_i$ label fundamental representations of each member of the pair. Then, we conjecture the following holomorphic relation:
\be
\begin{split}
  \ev{J_{A_1} J_{A_2} J_{A_3} J_{A_4}} (z)
  &=\sum_\alpha \cF_{\alpha,a_1\ba_2\ba_3a_4}(z)\,\tcF_{\alpha,\ta_1\bta_2\bta_3\ta_4}(z)\\
  &=
  \sum_\alpha\sum_{p,\tp}D^{(p)}_{a_1\ba_2\ba_3a_4}\tD^{(\tp)}_{\ta_1\bta_2\bta_3\ta_4}\, \cF_\alpha^{(p)}(z)\tcF_\alpha^{(\tp)}(z)
\end{split}
\label{cosetconj}
\ee

\section{Testing the Bilinear Relation for Conformal Blocks}

\label{testingsec}

We now test the conjecture of \eref{cosetconj} in all the possible cases.

\subsection{A$_{1,1} \oplus$ E$_{7,1}$}

In this case we choose the factor $(\mathbf{2},\mathbf{56})$ in the decomposition of the adjoint index of E$_{8,1}$ in the first line of \eref{decomp}. Thus adjoint indices $A_i$  of E$_{8,1}$ are treated as composite indices $(a_i\ta_i)$ where $a_i=1,2$ and $\ta_i=1,2,\cdots,56$. These are both pseudo-real representations, which means the complex conjugates of the fields $g_{ab}$ are related to the original fields by:
\be
{\bar g}_{\ba\bb}=\epsilon_{\ba \ba'}\,\epsilon_{\bb\bb'}\,g_{a'b'}
\ee
and a sum over repeated indices (one unbarred and one barred) is implied. Here for A$_1$, $\epsilon$ is the standard antisymmetric $\epsilon$-symbol, while for E$_7$ it is the symplectic invariant that we will define below (where it will be denoted $\tepsilon_{\ta\tb}$). Using this one can go back and forth between the correlator of two unbarred and two barred fields, and four unbarred fields. As indicated earlier, for notational simplicity we will make the latter choice i.e. four copies of the same (fundamental) field. 

For the A$_{1,1}$ WZW model we have $c = 1$ and $h = \frac14$. 
There is only one conformal block for the four-point function, $\alpha=1$, as the second one decouples (this is a general feature of A$_{n,1}$). Correspondingly the $D^{(p)}_{a_1a_2a_3a_4}$ are tensors that combine $\mathbf 2\times \mathbf 2$ into $\mathbf 3$ (symmetric) and $\mathbf 1$ (antisymmetric) respectively, in the $(a_1a_4)$ and $(a_2a_3)$ channels. Explicitly, they are:
\be
\begin{split}
D^{(0)}_{a_1a_2a_3a_4} &=\shalf(\epsilon_{a_1a_2}\epsilon_{a_3a_4}+\epsilon_{a_1 a_3}\epsilon_{a_2a_4})\\
D^{(1)}_{a_1a_2a_3a_4} &=\shalf(\epsilon_{a_1a_2}\epsilon_{a_3 a_4}-\epsilon_{a_1a_3}\epsilon_{a_2 a_4})=-\shalf\epsilon_{a_1a_4}\epsilon_{a_2a_3}
\end{split}
\label{Dforsu2}
\ee
The identity conformal block is given in the two cases by \cite{Knizhnik:1984nr}:
\begin{equation}
  \begin{split}
\cF_1^{(0)}(z) &= (z(1-z))^{\half}\left(\frac{1}{z}+\frac{1}{1-z}\right)\\
   \cF_1^{(1)}(z) &= (z(1-z))^{\half}\left(\frac{1}{z}-\frac{1}{1-z}\right)
  \end{split}
\end{equation}
As one can see, these sub-blocks have the same symmetry under $z\to 1-z$ as the tensor structures they multiply. The complete block is:
\be
\cF_{1}^{{\rm A}_{1,1}}(z)=\sum_{p=0}^1 \cF_1^{(p)}(z)D^{(p)}=
(z(1-z))^{\half}\left(
\frac{(D^{(0)}+D^{(1)})}{z}+
\frac{(D^{(0)}-D^{(1)})}{1-z}\right)
\label{su2block}
\ee
We have suppressed the indices on both sides to avoid clutter.

The E$_{7,1}$ WZW model has $c=7$ and $h_A=\frac34$.
We need the conformal blocks for the correlator $\ev{\mathbf{56}~\mathbf{56}~\mathbf{56}~\mathbf{56}}$. For this, we first note that in terms of representations of the E$_7$ Lie algebra, the tensor product of the fundamental with itself is:
\begin{equation}
  \mathbf{56}\otimes \mathbf{56}= \mathbf{1}\oplus\mathbf{133}\oplus \mathbf{1463}\oplus   \mathbf{1539}\\
\end{equation}
It is easily verified that the $\mathbf{133}$ and the $\mathbf{1463}$ are contained in the symmetric part of the product, while the $\mathbf{1}$ and the $\mathbf{1539}$ are in the antisymmetric part. At level 1 the  $\mathbf{133},\mathbf{1463}$ and $\mathbf{1539}$ all decouple, leaving the $\mathbf{56}$ as the only non-trivial primary.  Thus there is one block, just as for A$_{1,1}$.

In \cite{Mukhi:2017ugw} it was argued that in this case the identity and its descendants up to level 3 can flow in different sub-blocks in the $(14)$ fusion channel. Hence, for consistency there must be precisely four invariant tensors of E$_{7,1}$ contributing to the block, and these should be in correspondence with the four representations  $\mathbf{1}$, $\mathbf{133}$, $\mathbf{1463}$ and $\mathbf{1539}$. We will show below that this is indeed the case. 

As explained in Section \ref{WZWrev}, the invariant tensors are labelled\footnote{Recall that in a coset pair, all quantities pertaining to the second member are denoted with a tilde.} $\tD^{(p)}_{\ta_1\bta_2\bta_3\ta_4}$ with $p=0,1,2,3$ and each one will be part of either $D^S$ or $D^A$ defined in \eref{DSDA} according to its symmetry under $z\to 1-z$. Of these, $\tD^{(3)}$ corresponds to fusion into the primary, which in this case is the identity. The sub-blocks corresponding to these tensor structures were computed in \cite{Mukhi:2017ugw}:
\begin{equation}
  \begin{split}
    \tcF_1^{(0)}(z) &= (z(1-z))^{-\frac12}\left(\frac{1}{z}+\frac{1}{1-z}\right) \\
    \tcF_1^{(1)}(z) &= (z(1-z))^{-\half}\left(\frac{1}{z}-\frac{1}{1-z}\right)  \\
    \tcF_1^{(2)}(z) &= (z(1-z))^{-\half}\left(\frac{1}{z}+\frac{1}{1-z}+2\right) \\
    \tcF_1^{(3)}(z) &= (z(1-z))^{-\half} \left(\frac{1}{z}-\frac{1}{1-z}+14(1-2z)\right) 
  \end{split}
  \label{e7subblocks}
\end{equation}

The next step is to identify the tensor structures corresponding to these blocks. As explained in \cite{Mukhi:2017ugw}, the highest allowed value of $p$ corresponds to the flow of the primary in the $(14)$ channel, while other values correspond to secondaries. In the present case this means that the sub-block for $p=3$ corresponds to the identity, which appears in the antisymmetric part of $\mathbf{56}\otimes \mathbf{56}$. Therefore $\tD^{(3)}$ must be antisymmetric in $\ta_1,\ta_4$ and $\ta_2,\ta_3$.
The level of the secondary associated with the rest of the sub-blocks is $3-p$. We need to identify which representation corresponds to each of these secondary levels.
To get the first level descendant, we have to act with $J^a_{-1}$ on the primary.
Since the currents are in the adjoint representation, we obtain $\mathbf{133}$ at level 1, this corresponds to $\tD^{(2)}$ and is included in the symmetric part of $\mathbf{56}\otimes \mathbf{56}$. At the next descendant level, we act with $J^a_{-1}$ on $\mathbf{133}$.
In the decomposition of the tensor product of $\mathbf{133}\otimes \mathbf{133}$, we find the representation $\mathbf{1539}$. Thus the second level descendant contains $\mathbf{1539}$, which corresponds to the antisymmetric tensor structure $\tD^{(1)}$.
Finally for the third-level descendant we  consider $\mathbf{133}\otimes\mathbf{1539}$ where we find the representation $\mathbf{1463}$, corresponding to the symmetric invariant $\tD^{(0)}$. 

Thus we have argued that the tensors $\tD^{(0)}$ and $\tD^{(2)}$ are symmetric and $\tD^{(1)}$ and $\tD^{(3)}$ are antisymmetric in the indices $(\ta_1, \ta_4)$ and $(\ta_2,\ta_3)$. The associated representations flowing in each case are:
\be
\begin{split}
\tD^{(0)}\!\!: &\qquad \mathbf{1463}~~~~~{\rm (S)}\\  
\tD^{(1)}\!\!: &\qquad \mathbf{1539}~~~~~{\rm (A)}\\  
\tD^{(2)}\!\!: &\qquad \mathbf{~~133}~~~~~{\rm (S)}\\  
\tD^{(3)}\!\!: &\qquad ~~~~~\,\mathbf{1}~~~~~{\rm (A)}
\end{split}
\ee
This is confirmed by the fact that out of the sub-blocks in \eqref{e7subblocks}, $\tcF_1^{(0)}$ and $\tcF_1^{(2)}$ are indeed symmetric while $\tcF_1^{(1)}$ and $\tcF_1^{(3)}$ are antisymmetric under the exchange $z\leftrightarrow 1-z$. This is an important test of the formulae in \cite{Mukhi:2017ugw}. 

Our analysis in particular tells us that:
\be
\begin{split}
(\tD^{(0)}+\tD^{(2)})_{\ta_1\ta_2\ta_3\ta_4}
 &= \shalf(\tepsilon_{\ta_1\ta_2}\tepsilon_{\ta_3\ta_4}+\tepsilon_{\ta_1\ta_3}\tepsilon_{\ta_2\ta_4})\\
(\tD^{(1)}+\tD^{(3)})_{\ta_1\ta_2\ta_3\ta_4} &= \shalf(\tepsilon_{\ta_1\ta_2}\tepsilon_{\ta_3\ta_4}-\tepsilon_{\ta_1\ta_3}\tepsilon_{\ta_2\ta_4})
\end{split}
\label{e7Dsums}
\ee
Thus if we know one of $\tD^{(0)}, \tD^{(2)}$ and one of $\tD^{(1)},\tD^{(3)}$ then we can find all the four $\tD^{(p)}$ using the above equation. So it is sufficient to find, say, $\tD^{(3)}$ and $\tD^{(2)}$ which are, respectively, antisymmetric and symmetric under $\ta_1\leftrightarrow\ta_4$. For the antisymmetric case, we note that the E$_7$ symplectic tensor:
\be
\tepsilon_{\ta\tb}=\begin{pmatrix} 
\mathbf{0} & \mathbf{1_{28}}\\
-\mathbf{1_{28}} & \mathbf{0}
\end{pmatrix}
\ee
serves to combine two fundamentals into the identity. This is precisely what contributes to the invariant $\tD^{(3)}$, so we can write:
\be
\tD^{(3)}_{\ta_1\ta_2\ta_3\ta_4} \sim \tepsilon_{\ta_1\ta_4}\tepsilon_{\ta_2\ta_3}
\label{D3.E7}
\ee
where the normalisation remains to be determined.

Notice that unlike the A$_{1,1}$ case, here there is no identity relating this invariant to 
$\tepsilon_{\ta_1\ta_2}\tepsilon_{\ta_3\ta_4}-\tepsilon_{\ta_1\ta_3}\tepsilon_{\ta_2\ta_4}$.
This is a good thing because as explained above we need two independent invariants that are antisymmetric under $a_1\leftrightarrow a_4$. 

To find the symmetric invariant $\tD^{(2)}$ we use the third-rank tensor that combines two $\mathbf{56}$'s into the $\mathbf{133}$. Denote it by $\tq_{\ta\tb \tmu}$ where $\ta,\tb\in 1,\cdots,56$ and $\tmu\in 1,2,\cdots,133$. Then we can write:
\be
\tD^{(2)}_{\ta_1\ta_2\ta_3\ta_4}=\tq_{\ta_1\ta_4 \tmu}\,\tq_{\ta_2\ta_3 \tmu}
\label{D2.E7}
\ee
where $\tmu$ is summed over. We have absorbed a possible normalisation factor into the definition of $\tq_{\ta\tb\mu}$. We now have all four tensor structures $\tD^{(p)}$.

Combining the sub-blocks in \eref{e7subblocks} with their corresponding tensor structures, we can write the full E$_{7,1}$ block:
\begin{equation}
  \begin{split}
    \tcF_1^{{\rm E}_{7,1}} (z) &= \sum_{p=0}^3  \tcF^{(p)}_1 \tD^{(p)}\\
    &=(z(1-z))^{-\half}\Bigg[ \left(\tD^{(0)}+\tD^{(1)}+\tD^{(2)}+\tD^{(3)}\right)\frac{1}{z}\\[2mm]
      &\qquad + \left(\tD^{(0)}-\tD^{(1)}+\tD^{(2)}-\tD^{(3)}\right)\frac{1}{1-z} + 2\tD^{(2)}+14(1-2z)\tD^{(3)}\Bigg]
  \end{split}
  \label{e7block}
\end{equation}

Now we can investigate whether the conjectured coset relation \eref{cosetconj} holds.
Multiplying \eref{su2block} and \eref{e7block}, we find:
\begin{equation}
  \begin{split}
 &   \cF_1^{{\rm A}_{1,1}}(z) \, \tcF_1^{{\rm E}_{7,1}} (z) = 
    \frac{(D^{(0)}+D^{(1)})(\tD^{(0)} + \tD^{(1)} + \tD^{(2)} + \tD^{(3)})}{z^2} \\
    &\qquad +\frac{(D^{(0)}-D^{(1)})(\tD^{(0)} - \tD^{(1)} + \tD^{(2)} - \tD^{(3)})}{(1-z)^2} - 56 D^{(1)} \tD^{(3)} \\
    &\qquad + \frac{2(D^{(0)} \tD^{(0)} - D^{(1)} \tD^{(1)} + D^{(0)} \tD^{(2)} - D^{(1)} \tD^{(3)}) + (D^{(0)}+D^{(1)}) (2 \tD^{(2)} + 14 \tD^{(3)})}{z} \\
    &\qquad + \frac{2(D^{(0)} \tD^{(0)} - D^{(1)} \tD^{(1)} + D^{(0)} \tD^{(2)} - D^{(1)} \tD^{(3)}) + (D^{(0)}-D^{(1)}) (2 \tD^{(2)} - 14 \tD^{(3)})}{1-z}
  \end{split}
\label{su2e7prod}
\end{equation}
This is to be compared with the E$_{8,1}$ current correlator \eref{currentcorr} specialised to the relevant composite indices in A$_{1,1}$ and E$_{7,1}$. We already see considerable evidence in support of the conjectured relation. \eref{su2e7prod}, unlike the individual correlators for $A_{1,1}$ and $E_{7,1}$, has single and double poles at $z=0,z=1$ and a constant term -- precisely the structure of current four-point functions in \eref{currentcorr}. 

To compare in more detail, we must make the following reductions in \eref{currentcorr}:
\be
\begin{split}
  \delta_{A_1A_2}\delta_{A_3A_4}&\to \epsilon_{a_1a_2}\epsilon_{a_3a_4} \tepsilon_{\ta_1\ta_2}\tepsilon_{\ta_3\ta_4}\\
  \delta_{A_1A_3}\delta_{A_2A_4}&\to \epsilon_{a_1a_3}\epsilon_{a_2a_4} \tepsilon_{\ta_1\ta_3}\tepsilon_{\ta_2\ta_4}\\
  \delta_{A_1A_4}\delta_{A_2A_3}&\to \epsilon_{a_1a_4}\epsilon_{a_2a_3} \tepsilon_{\ta_1\ta_4}\tepsilon_{\ta_2\ta_3}
\end{split}
\label{e8su2e7decomp}
\ee
As mentioned earlier, $\epsilon_{ab}$ is the antisymmetric tensor in two dimensions, while $\tepsilon_{\ta\tb}$ is the symplectic E$_{7,1}$ invariant. 

Examining the double-pole terms\footnote{The constant term is also a ``double-pole" term if we consider the behaviour at infinity.} of \eref{su2e7prod} and using \eref{e8su2e7decomp}, we find that for the conjecture to be true, the following identities must hold:
\be
\begin{split}
  D^{(0)}+D^{(1)}&=p\, \epsilon_{a_1a_2}\epsilon_{a_3a_4}, \quad 
  \tD^{(0)}+\tD^{(1)}+\tD^{(2)}+\tD^{(3)} =p^{-1}\tepsilon_{\ta_1\ta_2}\tepsilon_{\ta_3\ta_4}\\
  D^{(0)}-D^{(1)}&= q\,\epsilon_{a_1a_3}\epsilon_{a_2a_4}, \quad 
  \tD^{(0)}-\tD^{(1)}+\tD^{(2)}-\tD^{(3)} =q^{-1}\tepsilon_{\ta_1\ta_3}\tepsilon_{\ta_2\ta_4}\\
  D^{(1)} &=r\, \epsilon_{a_1a_4}\epsilon_{a_2a_3},\quad 
  \tD^{(3)} =- \frac{r^{-1}}{56}\, \tepsilon_{\ta_1\ta_4}\tepsilon_{\ta_2\ta_3}
\end{split}
\label{su2e7tensors}
\ee
where $p,q,r$ are possible proportionality factors. Before verifying these relations, let us notice that the proposal has already passed one more test. The invariant $\tD^{(2)}$ in \eref{D2.E7} cannot appear on the RHS of \eref{e8su2e7decomp} because the LHS is just a product of Kronecker $\delta$'s which can only reduce to $\delta$ and $\epsilon$ symbols in the subgroups. Fortunately it also does not appear in the first two lines of \eref{su2e7tensors}, except in combination with $\tD^{(0)}$ into a product of $\tepsilon$'s by \eref{e7Dsums}. 

Consulting Eq.(\ref{Dforsu2}) we see that the above relations for $D^{(p)}$ are true and they determine $p=1,q=1,r=-\half$. From Eqs.(\ref{e7Dsums}, \ref{D3.E7}) we 
conclude that the four tensor structures arising in the E$_{7,1}$ 4-point function are:
\begin{equation}
  \begin{split}
    \tD^{(0)}_{\,\ta_1\ta_2\ta_3\ta_4} &= \shalf (\tepsilon_{\ta_1 \ta_2}\tepsilon_{\ta_3\ta_4} + 
    \tepsilon_{\ta_1 \ta_3}\tepsilon_{\ta_2\ta_4}) - \tD^{(2)}_{\,\ta_1\ta_2\ta_3\ta_4}\\
    \tD^{(1)}_{\,\ta_1\ta_2\ta_3\ta_4} &= \shalf  (\tepsilon_{\ta_1 \ta_2}\tepsilon_{\ta_3\ta_4} - 
    \tepsilon_{\ta_1 \ta_3}\tepsilon_{\ta_2\ta_4}) - \tD^{(3)}_{\,\ta_1\ta_2\ta_3\ta_4}\\
    \tD^{(2)}_{\,\ta_1\ta_2\ta_3\ta_4} &=\tq_{\ta_1\ta_4\tmu}\,\tq_{\ta_2\ta_3\tmu}\\
    \tD^{(3)}_{\,\ta_1\ta_2\ta_3\ta_4} &= \sfrac{1}{28}\tepsilon_{\ta_1\ta_4}\tepsilon_{\ta_2 \ta_3} 
  \end{split}
\end{equation}
Thus we have been able to fix all the tensor structures and also the normalisation of $\tD^{(3)}$ (but not the normalisation of $\tD^{(2)}$, since $\tq_{\ta\tb\tmu}$ has not yet been normalised) just by looking at the double-pole terms in the current correlator \eref{currentcorr}.

Let us now turn to the single-pole terms, which depend on the structure constants. 
For the conjecture to be true the pole at $z=0$ must match, which gives:
\begin{align}
  \begin{split}
    f_{A_1 A_2 B} f_{B A_3 A_4}&= 
    \eps_{a_1 a_2}\eps_{a_3 a_4}\wtd\eps_{\ta_1 \ta_3}\wtd\eps_{\ta_2 \ta_4}
    +\eps_{a_1 a_3}\eps_{a_2 a_4}\wtd\eps_{\ta_1 \ta_2}\wtd\eps_{\ta_3 \ta_4}
    \\
    &\quad
    +\sfrac12\eps_{a_1 a_2}\eps_{a_3 a_4}\wtd\eps_{\ta_1 \ta_4}\wtd\eps_{\ta_2 \ta_3}
    + 2\eps_{a_1 a_2}\eps_{a_3 a_4} \tq_{\ta_1 \ta_4 \tmu}\, \tq_{\ta_2 \ta_3\tmu}
  \end{split}\label{sglpole0}
\end{align}
The symmetry of both \eref{currentcorr} and \eref{su2e7prod} under $2\leftrightarrow 3$ together with $z\to 1-z$ ensures that the pole at $z=1$ gives the same answer after a re-labeling. 

The LHS of \eref{sglpole0} is separately antisymmetric under $A_1\leftrightarrow A_2$ and $A_3\leftrightarrow A_4$. Enforcing the corresponding antisymmetry on the RHS, we get the following condition:
\begin{equation}
\tq_{\ta_1 \ta_3 \tmu}\,\tq_{\ta_4 \ta_2 \tmu}-\tq_{\ta_1 \ta_4 \tmu}\,\tq_{\ta_2 \ta_3 \tmu}
=\shalf\left(
  \wtd \eps_{\ta_1 \ta_2}\wtd \eps_{\ta_3 \ta_4} -\sfrac12 \wtd \eps_{\ta_1 \ta_3}\wtd \eps_{\ta_4 \ta_2}-\sfrac12 \wtd \eps_{\ta_1 \ta_4}\wtd \eps_{\ta_2 \ta_3}\right)
\label{asym1}
\end{equation}
This non-trivial relation between different tensor invariants in E$_7$ has been proved in the mathematics literature \cite{Adams:book}. It follows immediately from Theorem 12.6 of that reference, with the following identifications. $W$ in the reference is the $\mathbf{57}$ representation, while $A$ is the $\mathbf{133}$. The skew-symmetric bilinear form on $W$, denoted by $\langle~,~\rangle$, is just our symplectic invariant $\tepsilon$. The map $\circ: W\otimes W\to A$ is proportional to our $\tq_{\ta\tb\tmu}$. Finally the inner product $(~,~)$ on $A$ is proportional to the Killing form, however the latter is negative definite for compact simple Lie algebras so it has to be proportional, with a real factor, to {\em minus} the sum over $\tmu$ for us. Thus we must take:
\be
(w\circ x, y\circ z)=-X\,\tq_{\ta_w\ta_x\tmu}\,\tq_{\ta_y\ta_z\tmu}
\ee
where $X$ is a normalisation to be determined.

Now from Theorem 12.6 of \cite{Adams:book} and the definition above it, one can easily prove the identity:
\be
(w\circ x, y\circ z)-(w\circ z,x\circ y)=
\langle w,y\rangle \langle z,x\rangle -\shalf \langle w,z\rangle \langle x,y\rangle - \shalf \langle w,x\rangle \langle y,z\rangle
\ee
which after transcribing to our notation and with the substitution $w\to a_1, x\to a_4, y\to a_2, z\to a_3$, becomes:
\be
-X (\tq_{\ta_1\ta_4\tmu}\,\tq_{\ta_2\ta_3\tmu}-
\tq_{\ta_1\ta_3\tmu}\,\tq_{\ta_4\ta_2\tmu})
=
\tepsilon_{\ta_1\ta_2}\tepsilon_{\ta_3\ta_4}
-\shalf
\tepsilon_{\ta_1\ta_3}\tepsilon_{\ta_4\ta_2}
-\shalf
\tepsilon_{\ta_1\ta_4}\tepsilon_{\ta_2\ta_3}
\label{adamsthm}
\ee
It is easily seen that \eref{adamsthm} is the same as \eref{asym1} with the choice $X=2$, i.e. the correct normalisation is $w\circ x\to \sqrt2\, \tq$. 

To summarise, the desired antisymmetry of the RHS of \eref{sglpole0} has been proved with the help of a very non-trivial E$_7$ identity! We can now check if it satisfies the Jacobi identity, as it must if the conjectured relation is to be true. Cyclically permuting the indices of \eqref{sglpole0} and adding, the RHS gives:
\begin{equation}
  \begin{split}
    &\eps_{a_1 a_2} \eps_{a_3 a_4} \left(\wtd \eps_{\ta_1 \ta_2} \wtd \eps_{\ta_3 \ta_4} -\sfrac12 \wtd \eps_{\ta_1 \ta_3}\wtd \eps_{\ta_4 \ta_2} -\sfrac12\wtd \eps_{\ta_1 \ta_4} \wtd \eps_{\ta_2 \ta_3} +2(\tq_{\ta_1 \ta_4 \mu}\, \tq_{\ta_2 \ta_3 \mu} - \tq_{\ta_1 \ta_3 \mu}\,\tq_{\ta_4 \ta_2 \mu})\right) \\[2mm]
    &+\eps_{a_1 a_3}\eps_{a_4 a_2} \left(-\wtd \eps_{\ta_1 \ta_4} \wtd \eps_{\ta_2 \ta_3} +\sfrac12\wtd \eps_{\ta_1 \ta_2}\wtd \eps_{\ta_3 \ta_4} +\sfrac12\wtd \eps_{\ta_1 \ta_3} \wtd \eps_{\ta_4 \ta_2} +2 (\tq_{\ta_1 \ta_2 \mu}\, \tq_{\ta_3 \ta_4 \mu} - \tq_{\ta_1 \ta_3 \mu}\,\tq_{\ta_4 \ta_2 \mu})\right) = 0 
  \end{split}
\end{equation}
This is indeed seen to vanish using \eref{asym1}. 

Thus we see that \eref{asym1}, which is a quadratic relation between the tensor invariants $\tepsilon_{\ta\tb}$ and $\tq_{\ta\tb\mu}$ of E$_7$, is sufficient to ensure that the RHS of \eref{sglpole0} has the same symmetries as the LHS of that equation -- namely, antisymmetry in a pair of indices as well as the Jacobi identity. Together with the fact that the overall pole structure matches perfectly, this amounts to strong confirmation of the correctness of the conjecture \eref{cosetconj} for the present case.

Let us mention that in this particular example, we would recover the identity \eref{asym1} just by requiring crossing invariance of the single conformal block (up to a phase). However in subsequent examples, particularly those with complex fields, we have to implement crossing by summing over all orderings of the fields. In those cases we will not need identities among tensor invariants of the algebra.

\subsection{A$_{2,1}\oplus$ E$_{6,1}$}

For this case, we focus on the factor $(\mathbf{3},\mathbf{27})+(\mathbf{\overline{3}},\mathbf{\overline{27}})$ in the decomposition of the adjoint index of E$_{8,1}$ as in the second line of \eref{decomp}. Thus, adjoint indices $A_i$ of E$_{8,1}$ will now be composite holomorphic indices $(a_i\ta_i)$ where $a_i=1,2,3$ and $\ta_i=1,2,\cdots,27$, or else anti-holomorphic indices $(\ba_i\bta_i)$. We must sum over both. As a result we get a total of 16 terms, of which 10 vanish because the number of holomorphic and anti-holomorphic fields is different. The remaining 6 terms have the form:
\be
\begin{split}
&\Big(\langle \mathbf{3}\, \mathbf{\overline{3}}\,\mathbf{\overline{3}}\,\mathbf{3} \rangle_{a_1\ba_2\ba_3 a_4} 
\langle \mathbf{27} \,\mathbf{\overline{27}}\,\mathbf{\overline{27}}\,\mathbf{27} \rangle_{\ta_1\bta_2\bta_3 \ta_4}
 + 
 \langle \mathbf{3}\, \mathbf{{3}}\,\mathbf{\overline{3}}\, \mathbf{\overline{3}}\rangle_{a_1a_2\ba_3 \ba_4}
 \langle \mathbf{27}\, \mathbf{27}\,\mathbf{\overline{27}}\, \mathbf{\overline{27}}\rangle_{\ta_1\ta_2\bta_3 \bta_4} \\
&\quad + \langle \mathbf{3}\, \mathbf{\overline{3}}\,\mathbf{3}\,\mathbf{\overline{3}}\rangle_{a_1\ba_2 a_3 \ba_4}
\langle\mathbf{27}\,\mathbf{\overline{27}}\,\mathbf{27}
\,\mathbf{\overline{27}}
\rangle_{\ta_1\bta_2\ta_3 \bta_4}
\Big)+\hbox{cc}
\end{split}
\label{compcrossing}
\ee
The conformal blocks appearing in each term above are related to those in other terms by crossing. In this particular case there is just one block (with two sub-blocks) for each of the theories in the coset pair, hence crossing must bring a block back to itself up to a possible phase (more generally it sends the blocks to linear combinations of themselves). However even in this simple case, the functional form of the block in the limit $z_2,z_3,z_4\to 0,1,\infty$ is different for each of the six ways of choosing two fundamental and two anti-fundamental representations in the four-point functions. Hence we will start with the first term above and then deduce the other two terms by making conformal transformations $z\to \frac{1}{z}$ and $z\to \frac{z}{z-1}$ which respectively interchange 0 with $\infty$ and 1 with $\infty$. 

For A$_{2,1}$, we have $c = 2$ and $h = \frac13$. The correlator $\ev{\mathbf{3}\,\ovl{\mathbf{3}}\,\ovl{\mathbf{3}}\,\mathbf{3}}$ gets a contribution from a single conformal block, as noted above. In the 14 channel, this corresponds to the conformal family of the $\mathbf{\ovl 3}$. The second block decouples at level 1. There are two sub-blocks, corresponding to the fusion of $\mathbf{3}\times \mathbf{3}$ into the $\mathbf{6}$ (symmetric) and $\mathbf{\bar 3}$ (antisymmetric) representations. The invariant tensors are:
\be
\begin{split}
D^{(0)}_{a_1\ba_2\ba_3 a_4} &= \shalf(\delta_{a_1\ba_2}\delta_{\ba_3 a_4}+
\delta_{a_1\ba_3}\delta_{\ba_2 a_4})\\
D^{(1)}_{a_1\ba_2\ba_3 a_4} &= \shalf(\delta_{a_1\ba_2}\delta_{\ba_3 a_4}-
\delta_{a_1\ba_3}\delta_{\ba_2 a_4})
\end{split}
\label{D0D1.SU3}
\ee
The corresponding sub-blocks are:
\begin{equation}
  \begin{split}
    \cF_1^{(0)} &= (z(1-z))^{\frac13} \left(\frac1z+\frac{1}{1-z}\right)\\
    \cF_1^{(1)} &= (z(1-z))^{\frac13} \left(\frac1z-\frac{1}{1-z}\right)
  \end{split}
\end{equation}
and the complete A$_{2,1}$ block is:
\be
\begin{split}
 \langle \mathbf{3}\, \mathbf{\overline{3}}\,\mathbf{\overline{3}}\,\mathbf{3} \rangle_{a_1\ba_2\ba_3 a_4} 
  &= \sum_{p=0}^1 \cF_1^{(p)} D^{(p)}_{a_1\ba_2\ba_3 a_4}\\
  &=
  (z(1-z))^{\frac13} \left(\frac{\delta_{a_1\ba_2}\delta_{\ba_3 a_4}}{z} + \frac{\delta_{a_1\ba_3}\delta_{\ba_2 a_4}}{1-z}\right)
 \label{A1.fullblock}
\end{split}
\ee
This time we have written out the indices explicitly because they will be important. We can identify $D^{(1)}$ with the flow of the $\ovl{\mathbf{3}}$ and $D^{(0)}$ with the $\mathbf{6}$.

Now applying the conformal transformations $z\to \frac{1}{z}$ and $\frac{z}{z-1}$ respectively, we get:
\be
\begin{split}
\ev{\mathbf{3}\,\mathbf{3}\,\overline{\mathbf{3}}\,\overline{\mathbf{3}}}_{a_1 a_2\ba_3\ba_4}
&:\left(\frac{1-z}{z}\right)^{\frac13}\left(\delta_{a_1 \ba_4}\delta_{\ba_3 a_2}-\frac{\delta_{a_1 \ba_3} \delta_{a_2 \ba_4}}{1-z}\right)\\   
    \ev{\mathbf{3}\,\overline{\mathbf{3}}\,\mathbf{3}\,\overline{\mathbf{3}}}_{a_1 \ba_2 a_3\ba_4}
    &:\left(\frac{z}{1-z}\right)^{\frac13}\left(\frac{\delta_{a_1 \ba_2}\delta_{a_3 \ba_4}}{z}-\delta_{a_1 \ba_4} \delta_{\ba_2 a_3}\right)
\end{split}
\ee

Next we turn to E$_{6,1}$, for which we have $c = 6$ and $h=\frac23$ and start with the correlator $\ev{\mathbf{27}\,\ovl{\mathbf{27}}\,\ovl{\mathbf{27}}\,\mathbf{27}}$. As in the previous case, the second block decouples
at level 1. The tensor product of the E$_6$ fundamental with itself decomposes as:
\begin{equation}
  \mathbf{27}\otimes\mathbf{27} = \ovl{\mathbf{27}} \oplus \ovl{\mathbf{351}} \oplus {\mathbf{\overline{351'}}}
\end{equation}
The symmetric part of $\mathbf{27}\otimes\mathbf{27}$ has total dimension 378. Thus it must contain the $\mathbf{\overline{27}}$ as well as one of the 
$\mathbf{\overline{351}}$ and $\mathbf{\overline{351'}}$.
 Since the Dynkin labels of the $\mathbf{27}$ and the $\mathbf{\overline{351'}}$ are $(1,0,0,0,0,0)$ and $(2,0,0,0,0,0)$ respectively, it is clear that the latter  lies in the symmetric product of the former with itself. Meanwhile the $\mathbf{\overline{351}}$, with Dynkin labels 
$(0,0,0,1,0,0)$, makes up the antisymmetric part of $\mathbf{27}\otimes\mathbf{27}$.
Clearly the $\mathbf{\ovl{27}}$ appears at the primary level, and one can verify that the secondaries, generated by currents in the adjoint $\mathbf{78}$, produce the representation $\mathbf{\ovl{351}}$ at the first level and 
$\mathbf{\ovl{351'}}$ at the second level. 

Thus we have:
\be
\begin{split}
\tD^{(0)}\!\!: &\qquad \mathbf{\ovl{351'}}~~~~~{\rm (S)}\\  
\tD^{(1)}\!\!: &\qquad \mathbf{\ovl{351}}~~~~~{\rm (A)}\\  
\tD^{(2)}\!\!: &\qquad \mathbf{~~\ovl{27}}~~~~~{\rm (S)}
\end{split}
\ee
We can now seek the corresponding tensor invariants. By the symmetry/antisymmetry argument above, we have:
\be
\begin{split}
(\tD^{(0)}+\tD^{(2)})_{\ta_1\bta_2\bta_3\ta_4} &= \shalf(\delta_{\ta_1\bta_2}\delta_{\bta_3\ta_4}+\delta_{\ta_1\bta_3}\delta_{\bta_2\ta_4})\\
\tD^{(1)}_{\ta_1\bta_2\bta_3\ta_4} &= \shalf(\delta_{\ta_1\bta_2}\delta_{\bta_3\ta_4}-\delta_{\ta_1\bta_3}\delta_{\bta_2\ta_4})
\end{split}
\label{D0D1D2.E6}
\ee
Thus we only need to find $\tD^{(2)}$. For this, we note that there is a 3-index tensor invariant $\tq_{\ta_1\ta_2\ta_3}$ that maps three $\mathbf{27}$'s to the singlet. The quartic invariant made from this will be the one corresponding to the flow of a $\mathbf{\overline{27}}$ in the  $\mathbf{27}\otimes\mathbf{27}$ channel. Thus we can write:
\be
\tD^{(2)}_{\ta_1\bta_2\bta_3\ta_4}= \sfrac19\, \tq_{\ta_1\ta_4\tb}\,\btq_{\bta_2\bta_3\btb}
\ee
where a sum over $\tb$ is implied in the last term. The factor of $\frac19$ has been introduced to simplify subequent formulae, it does not affect anything else since the normalisation of the $\tq$ has not yet been specified. 

The corresponding sub-blocks are \cite{Mukhi:2017ugw}:
\begin{equation}
  \begin{split}
    \tcF_1^{(0)} &= (z(1-z))^{-\frac13} \left(\frac1z+\frac{1}{1-z}\right)\\
    \tcF_1^{(1)} &= (z(1-z))^{-\frac13} \left(\frac1z-\frac{1}{1-z}\right)\\
    \tcF_1^{(2)} &= (z(1-z))^{-\frac13} \left(\frac1z+\frac{1}{1-z}+9\right)
  \end{split}
\end{equation}
and the complete block is:
\begin{equation}
\begin{split}
\langle \mathbf{27} \,\mathbf{\overline{27}}\,\mathbf{\overline{27}}\,\mathbf{27} \rangle_{\ta_1\bta_2\bta_3 \ta_4} 
  &= \sum_{p=0}^2  \tcF_1^{(p)}\tD^{(p)}_{\ta_1\bta_2\bta_3 \ta_4} \\
  &=(z(1-z))^{-\frac13} \Bigg(
  \frac{\delta_{\ta_1\bta_2}\delta_{\bta_3\ta_4}}{z} + \frac{\delta_{\ta_1\bta_3}\delta_{\bta_2\ta_4}}{1-z}  +  \tq_{\ta_1\ta_4 \tb}\btq_{\bta_2\bta_3 \btb}\Bigg)
\end{split}
\end{equation}
As before, we now exchange positions using conformal transformations and get:
\be
\begin{split}
\ev{\mathbf{27}\,\mathbf{27}\,\overline{\mathbf{27}}\,\overline{\mathbf{27}}}_{\ta_1 \ta_2\bta_3\bta_4}
&: \left(\frac{1-z}{z}\right)^{-\frac13}
    \left(\delta_{\ta_1 \bta_4}\delta_{\bta_3 \ta_2}-\frac{\delta_{\ta_1 \bta_3} \delta_{\bta_4 \ta_2}}{1-z}+\frac{\tq_{\ta_1 \ta_2 \tb} \btq_{\bta_3 \bta_4 \btb}}{z}\right)\\
\ev{\mathbf{27}\,\overline{\mathbf{27}}\,\mathbf{27}\,\overline{\mathbf{27}}}_{\ta_1 \bta_2\ta_3\bta_4}&:
\left(\frac{z}{1-z}\right)^{-\frac13}
    \left(\frac{\delta_{\ta_1 \bta_2}\delta_{\ta_3 \bta_4}}{z}-\delta_{\ta_1 \bta_4} \delta_{\bta_2 \ta_3}-\frac{\tq_{\ta_1 \ta_3 \tb} \btq_{\bta_2 \bta_4 \btb}}{1-z}\right)
\end{split}
\ee

Combining them, we get:
\begin{equation}
  \begin{split}
    & \ev{\mathbf{3}\,\overline{\mathbf{3}}\,\overline{\mathbf{3}}\,\mathbf{3}} \ev{\mathbf{27}\,\overline{\mathbf{27}}\,\overline{\mathbf{27}}\,\mathbf{27}}
    + \ev{\mathbf{3}\,\mathbf{3}\,\overline{\mathbf{3}}\,\overline{\mathbf{3}}} \ev{\mathbf{27}\,\mathbf{27}\,\overline{\mathbf{27}}\,\overline{\mathbf{27}}}
    + \ev{\mathbf{3}\,\overline{\mathbf{3}}\,\mathbf{3}\,\overline{\mathbf{3}}} \ev{\mathbf{27}\,\overline{\mathbf{27}}\,\mathbf{27}\,\overline{\mathbf{27}}} \\[2mm]
    &=\frac{1}{z^2}\Big(\delta_{a_1 \ba_2}\delta_{\ba_3 a_4} \delta_{\ta_1 \bta_2}\delta_{\bta_3 \ta_4}+\delta_{a_1 \ba_2}\delta_{a_3 \ba_4} \delta_{\ta_1 \bta_2}\delta_{\ta_3 \bta_4}\Big)
    \\
    &
   \quad + \frac{1}{(1-z)^2}\Big(\delta_{a_1 \ba_3}\delta_{\ba_2 a_4} \delta_{\ta_1 \bta_3}\delta_{\bta_2 \ta_4}+ \delta_{a_1 \ba_3}\delta_{a_2 \ba_4} \delta_{\ta_1 \bta_3}\delta_{\ta_2 \bta_4}\Big)\\
    &\quad+\delta_{a_1 \ba_4}\delta_{a_2 \ba_3} \delta_{\ta_1 \bta_4}\delta_{\ta_2 \bta_3} + \delta_{a_1 \ba_4}\delta_{\ba_2 a_3} \delta_{\ta_1 \bta_4}\delta_{\bta_2 \ta_3}\\
    &\quad+\frac{1}{z}\Big(
      \delta_{a_1 \ba_2}\delta_{\ba_3 a_4} \delta_{\ta_1 \bta_3}\delta_{\bta_2 \ta_4} +
  \delta_{a_1 \ba_3}\delta_{\ba_2 a_4}\delta_{\ta_1 \bta_2}\delta_{\bta_3 \ta_4}    
-\delta_{a_1 \ba_2}\delta_{\ba_4 a_3}
\delta_{\ta_1 \bta_4}\delta_{\bta_2 \ta_3}
-\delta_{a_1 \ba_4}\delta_{\ba_2 a_3}\delta_{\ta_1 \bta_2}\delta_{\bta_4 \ta_3}\\
      &\qquad 
      +\delta_{a_1 \ba_2}\delta_{\ba_3 a_4} \tq_{\ta_1 \ta_4 \tb}\btq_{\bta_2 \bta_3 \btb}
      -\delta_{a_1 \ba_2}\delta_{\ba_4 a_3} \tq_{\ta_1 \ta_3 \tb}\btq_{\bta_2 \bta_4 \btb}
      -(\delta_{a_1 \ba_3}\delta_{\ba_4 a_2}-\delta_{a_1 \ba_4}\delta_{\ba_2 a_3})\tq_{\ta_1 \ta_2 \tb}\btq_{\bta_3 \bta_4 \btb}
      \Big)\\
    &\quad+\frac{1}{1-z}
 \Big(      \delta_{a_1 \ba_3}\delta_{\ba_2 a_4} \delta_{\ta_1 \bta_2}\delta_{\bta_3 \ta_4} +
  \delta_{a_1 \ba_2}\delta_{\ba_3 a_4}\delta_{\ta_1 \bta_3}\delta_{\bta_2 \ta_4}    
-\delta_{a_1 \ba_3}\delta_{\ba_4 a_2}
\delta_{\ta_1 \bta_4}\delta_{\bta_3 \ta_2}
-\delta_{a_1 \ba_4}\delta_{\ba_3 a_2}\delta_{\ta_1 \bta_3}\delta_{\bta_4 \ta_2}\\
      &\qquad 
      +\delta_{a_1 \ba_3}\delta_{\ba_2 a_4} \tq_{\ta_1 \ta_4 \tb}\btq_{\bta_3 \bta_2 \btb}
      -\delta_{a_1 \ba_3}\delta_{\ba_4 a_2} \tq_{\ta_1 \ta_2 \tb}\btq_{\bta_3 \bta_4 \btb}
      -(\delta_{a_1 \ba_2}\delta_{\ba_4 a_3}-\delta_{a_1 \ba_4}\delta_{\ba_3 a_2})\tq_{\ta_1 \ta_3 \tb}\btq_{\bta_2 \bta_4 \btb}
      \Big)
    \\
  \end{split}
\label{A2.E6}
\end{equation}
We have dropped the complex conjugate on both sides. If we can verify the above relation, where the ``1'' index is fixed to be holomorphic, then it will hold for the full version with complex conjugates added.

Again the above expresses passes some basic tests: unlike the original conformal blocks, these products have double and single poles at $z=0,1$ as well as a constant term. This matches  with the structure of current correlation functions, and we can move on to perform more detailed tests.

So far we have kept Kronecker $\delta$'s like $\delta_{a_1\ba_2}$ and $\delta_{a_2\ba_1}$ distinct, to indicate their origin from different terms in the sum on the left side of \eref{A2.E6}. However they are really equal to each other so we may identify them. This allows us to simplify \eref{A2.E6} to:
\begin{equation}
  \begin{split}
    & \ev{\mathbf{3}\,\overline{\mathbf{3}}\,\overline{\mathbf{3}}\,\mathbf{3}} \ev{\mathbf{27}\,\overline{\mathbf{27}}\,\overline{\mathbf{27}}\,\mathbf{27}}
    + \ev{\mathbf{3}\,\mathbf{3}\,\overline{\mathbf{3}}\,\overline{\mathbf{3}}} \ev{\mathbf{27}\,\mathbf{27}\,\overline{\mathbf{27}}\,\overline{\mathbf{27}}}
    + \ev{\mathbf{3}\,\overline{\mathbf{3}}\,\mathbf{3}\,\overline{\mathbf{3}}} \ev{\mathbf{27}\,\overline{\mathbf{27}}\,\mathbf{27}\,\overline{\mathbf{27}}} \\[2mm]
    &=\frac{2\,\delta_{a_1 \ba_2}\delta_{a_3 \ba_4} \delta_{\ta_1 \bta_2}\delta_{\ta_3 \bta_4}}{z^2}
    + \frac{2\,\delta_{a_1 \ba_3}\delta_{a_2 \ba_4} \delta_{\ta_1 \bta_3}\delta_{\ta_2 \bta_4}}{(1-z)^2}
+2\,\delta_{a_1 \ba_4}\delta_{a_2 \ba_3} \delta_{\ta_1 \bta_4}\delta_{\ta_2 \bta_3}\\ 
    &\quad+\frac{1}{z}\Big(
\delta_{a_1 \ba_2}\delta_{a_3 \ba_4}(\delta_{\ta_1 \bta_3}\delta_{\ta_2 \bta_4}-\delta_{\ta_1 \bta_4}\delta_{\ta_2 \bta_3}+\tq_{\ta_1 \ta_4 \tb}\btq_{\bta_2 \bta_3 \btb}-\tq_{\ta_1 \ta_3 \tb}\btq_{\bta_2 \bta_4 \btb})\\
&\qquad\qquad +(\delta_{a_1 \ba_3}\delta_{a_2 \ba_4}-
 \delta_{a_1 \ba_4}\delta_{a_2 \ba_3})
(\delta_{\ta_1 \bta_2}\delta_{\ta_3 \bta_4}
-\tq_{\ta_1 \ta_2 \tb}\btq_{\bta_3 \bta_4\btb})\Big)
\\ &\quad+\frac{1}{1-z}
\Big(
\delta_{a_1 \ba_3}\delta_{a_2 \ba_4}(\delta_{\ta_1 \bta_2}\delta_{\ta_3 \bta_4}-\delta_{\ta_1 \bta_4}\delta_{\ta_2 \bta_3}+\tq_{\ta_1 \ta_4 \tb}\btq_{\bta_2 \bta_3 \btb}-\tq_{\ta_1 \ta_2 \tb}\btq_{\bta_3 \bta_4 \btb})\\
&\qquad\qquad +(\delta_{a_1 \ba_2}\delta_{a_3 \ba_4}-
 \delta_{a_1 \ba_4}\delta_{a_2 \ba_3})
(\delta_{\ta_1 \bta_3}\delta_{\ta_2 \bta_4}
-\tq_{\ta_1 \ta_3 \tb}\btq_{\bta_2 \bta_4\btb})\Big)    
    \\
  \end{split}
\label{A2.E6.new}
\end{equation}

Now we can start by comparing the double-pole terms of \eref{currentcorr} with those of \eref{A2.E6}. This time the decomposition of E$_8$ indices is:
\be
\begin{split}
  \delta_{A_1A_2}\delta_{A_3A_4}&~\to~ 
  (\delta_{a_1\ba_2}\delta_{\ta_1\bta_2}
  +\delta_{\ba_1 a_2}\delta_{\bta_1\ta_2})
  (\delta_{a_3\ba_4}\delta_{\ta_3\bta_4}
+  \delta_{\ba_3 a_4}\delta_{\bta_3\ta_4})\\
&\quad =2\,\delta_{a_1\ba_2}\delta_{a_3\ba_4}\delta_{\ta_1\bta_2}
\delta_{\ta_3\bta_4} +{\rm c.c.}
\\
\delta_{A_1A_3}\delta_{A_2A_4}&\to
(\delta_{a_1\ba_3}\delta_{\ta_1\bta_3}
  +\delta_{\ba_1 a_3}\delta_{\bta_1\ta_3})
  (\delta_{a_2\ba_4}\delta_{\ta_2\bta_4}
+  \delta_{\ba_2 a_4}\delta_{\bta_2\ta_4})\\
&\quad = 2\,\delta_{a_1\ba_3}\delta_{a_2\ba_4}\delta_{\ta_1\bta_3}
\delta_{\ta_2\bta_4}+{\rm c.c.}
\\
\delta_{A_1A_4}\delta_{A_2A_3}&\to
(\delta_{a_1\ba_4}\delta_{\ta_1\bta_4}
  +\delta_{\ba_1 a_4}\delta_{\bta_1\ta_4})
  (\delta_{a_2\ba_3}\delta_{\ta_2\bta_3}
+  \delta_{\ba_2 a_3}\delta_{\bta_2\ta_3})  \\
&\quad = 2\,\delta_{a_1\ba_4}\delta_{a_2\ba_3}\delta_{\ta_1\bta_4}
\delta_{\ta_2\bta_3}+{\rm c.c.}
\end{split}
\label{E8.A2.E6decomp}
\ee
In each case, the answers have been re-expressed in terms of a part with a holomorphic ``1'' index plus its complex conjugate.  With this, we see perfect agreement between the $\frac{1}{z^2},\frac{1}{(1-z)^2}$ and constant terms in \eref{currentcorr} and \eref{A2.E6}.

Notice that the third-rank tensor $\tq_{\ta_1\ta_2\ta_3}$ of E$_6$ does not appear in the above checks, analogous to the fact that the invariant $\tq_{\ta_1\ta_2\tmu}$ of E$_7$ did not appear in the checks of the double-pole terms in the  A$_1 \oplus$ E$_7$ case. 

We now turn to the single-pole terms in \eref{currentcorr}, which depend on the E$_8$ structure constants. Matching the coefficient of $\frac{1}{z}$ in \eref{currentcorr} and \eref{A2.E6}, the coset relation will hold if:
\begin{equation}
  \begin{split}
    f_{A_1 A_2 B} f_{B A_3 A_4}&= 
  \delta_{a_1 \ba_2}\delta_{a_3 \ba_4}(\delta_{\ta_1 \bta_3}\delta_{\ta_2 \bta_4}-\delta_{\ta_1 \bta_4}\delta_{\ta_2 \bta_3}+\tq_{\ta_1 \ta_4 \tb}\btq_{\bta_2 \bta_3 \btb}-\tq_{\ta_1 \ta_3 \tb}\btq_{\bta_2 \bta_4 \btb})\\
&\qquad\qquad +(\delta_{a_1 \ba_3}\delta_{a_2 \ba_4}-
 \delta_{a_1 \ba_4}\delta_{a_2 \ba_3})
(\delta_{\ta_1 \bta_2}\delta_{\ta_3 \bta_4}
+\tq_{\ta_1 \ta_2 \tb}\btq_{\bta_3 \bta_4\btb})    
\end{split}
\ee    
The pole at $z=1$ gives the same information.

It is evident that the RHS has the desired antisymmetry under the exchange $1\leftrightarrow 2$ or $3\leftrightarrow 4$. Moreover, if we sum the above expression over cyclic permutations of 2,3,4 we immediately find that it vanishes, confirming that the Jacobi identity is satisfied.

\subsection{D$_{4,1}\oplus$ D$_{4,1}$}

There are three 8-dimensional representations of D$_4$, which we label $\mathbf{8_v},\mathbf{8_s},\mathbf{8_c}$ for vector, spinor, conjugate spinor, and they are related by triality. Because of this, they all appear symmetrically in \eref{decomp}. We choose all the fields in the correlator to be in the $\mathbf{8_v}$. Then the discussion becomes particularly simple as we can just reduce the E$_{8,1}$ current correlators to those of its D$_{8,1}$ subalgebra. Thereafter, the free-fermion descriptions of 
D$_{8,1}$ and D$_{4,1}$ are sufficient to establish the coset relation. However we will go ahead and work it out in parallel to the other cases. 

Due to the above choice, we work with the $(\mathbf{8_v},\mathbf{8_v})$ term in \eref{currentcorr}. Now we have:
\be
\mathbf{8_v}\otimes \mathbf{8_v}=\mathbf{35_v}\oplus \mathbf{28}\oplus \mathbf{1}
\ee
where the $\mathbf{28}$ is the adjoint and corresponds to the antisymmetric product, while the $\mathbf{35_v}$ and singlet appear in the symmetric product. 

Again, there is just one conformal block, with three sub-blocks that are found to be the following:
\be
\begin{split}
  f_1^{(0)} &=\left(\frac{1}{z}+\frac{1}{1-z}\right)\\
  f_1^{(1)} &=\left(\frac{1}{z}-\frac{1}{1-z}\right)\\
  f_1^{(2)} &=-1
\end{split}
\ee
and the corresponding tensor structures:
\be
\begin{split}
  D^{(0)} &= \shalf(\delta_{a_1a_2}\delta_{a_3a_4}
  +\delta_{a_1a_3}\delta_{a_2a_4})\\
  D^{(1)} &= \shalf(\delta_{a_1a_2}\delta_{a_3a_4}
  -\delta_{a_1a_3}\delta_{a_2a_4})\\
  D^{(2)} &= \delta_{a_1a_4}\delta_{a_2a_3}
\end{split}
\ee
Here $D^{(2)}$ manifestly corresponds to fusion of two $\mathbf{8_v}$'s into the identity, while $D^{(1)}$ corresponds to the flow of the secondary $\mathbf{28}$ in this channel and $D^{(2)}$ corresponds to the flow of the secondary $\mathbf{35_v}$ which indeed appears in $\mathbf{28}\otimes\mathbf{28}$.

Thus, the complete conformal block is:
\be
\cF^{{\rm D}_{4,1}}_{1,a_1a_2a_3a_4}(z)=\frac{\delta_{a_1a_2}\delta_{a_3a_4}}{z}
+\frac{\delta_{a_1a_3}\delta_{a_2a_4}}{1-z}-
\delta_{a_1a_4}\delta_{a_2a_3}
\ee
Since both members of the proposed coset pair are D$_{4,1}$, we immediately write out the content of our conjecture in this case:
\be
\begin{split}
  \ev{J_{A_1} J_{A_2} J_{A_3} J_{A_4}}(z)
  &=\left(\frac{\delta_{a_1a_2}\delta_{a_3a_4}}{z}
  +\frac{\delta_{a_1a_3}\delta_{a_2a_4}}{1-z}-
  \delta_{a_1a_4}\delta_{a_2a_3}\right)\\
  &\qquad \times
  \left(\frac{\delta_{\ta_1\ta_2}\delta_{\ta_3\ta_4}}{z}
  +\frac{\delta_{\ta_1\ta_3}\delta_{\ta_2\ta_4}}{1-z}-
  \delta_{\ta_1\ta_4}\delta_{\ta_2\ta_3}\right)\\[3mm]
  &= \frac{\delta_{a_1a_2}\delta_{a_3a_4} \delta_{\ta_1\ta_2}\delta_{\ta_3\ta_4}}{z^2}
  +\frac{\delta_{a_1a_3}\delta_{a_2a_4} \delta_{\ta_1\ta_3}\delta_{\ta_2\ta_4}}{(1-z)^2}\\
  &\qquad +\delta_{a_1a_4}\delta_{a_2a_3}
  \delta_{\ta_1\ta_4}\delta_{\ta_2\ta_3}\\[2mm]
  &\qquad +
  \frac{\delta_{a_1a_2}\delta_{a_3a_4}\delta_{\ta_1\ta_3}\delta_{\ta_2\ta_4}+\delta_{a_1a_3}\delta_{a_2a_4}
    \delta_{\ta_1\ta_2}\delta_{\ta_3\ta_4}}{z(1-z)}\\[2mm]
  &\qquad -\frac{\delta_{a_1a_2}\delta_{a_3a_4}
    \delta_{\ta_1\ta_4}\delta_{\ta_2\ta_3}+
    \delta_{a_1a_4}\delta_{a_2a_3}
    \delta_{\ta_1\ta_2}\delta_{\ta_3\ta_4}}{z}\\[2mm]
  &\qquad-\frac{\delta_{a_1a_3}\delta_{a_2a_4}
    \delta_{\ta_1\ta_4}\delta_{\ta_2\ta_3}+
    \delta_{a_1a_4}\delta_{a_2a_3}
    \delta_{\ta_1\ta_3}\delta_{\ta_3\ta_4}}{1-z}
\end{split}
\ee
Looking at the first three terms, they are equal to:
\be
\frac{\delta_{A_1A_2}\delta_{A_3A_4}}{z^2}
+ \frac{\delta_{A_1A_3}\delta_{A_2A_4}}{(1-z)^2}
+\delta_{A_1A_4}\delta_{A_2A_3}
\ee
This precisely reproduces the first three terms in \eref{currentcorr}.

The remaining terms can be written:
\be
\begin{split}
&\frac{\delta_{a_1a_2}\delta_{a_3a_4}(\delta_{\ta_1\ta_3}\delta_{\ta_2\ta_4}
-\delta_{\ta_1\ta_4}\delta_{\ta_2\ta_3})
+(\delta_{a_1a_3}\delta_{a_2a_4}-
\delta_{a_1a_4}\delta_{a_2a_3})\delta_{\ta_1\ta_2}\delta_{\ta_3\ta_4}}
{z}\\
&+\frac{\delta_{a_1a_3}\delta_{a_2a_4}(\delta_{\ta_1\ta_2}\delta_{\ta_3\ta_4}
-\delta_{\ta_1\ta_4}\delta_{\ta_2\ta_3})
+(\delta_{a_1a_2}\delta_{a_3a_4}-
\delta_{a_1a_4}\delta_{a_2a_3})\delta_{\ta_1\ta_3}\delta_{\ta_2\ta_4}}
{1-z}
\end{split}
\ee
These have the same form as the last two terms of \eref{currentcorr}, and will be identical to them if the following identity holds for $E_8$ structure constants with composite indices $A_i$ restricted to the $(8,8)$ representation of D$_4\,\oplus\,$D$_4$: 
\be
\begin{split}  
  f_{A_1A_2B} f_{BA_3A_4}&=\delta_{a_1a_2}\delta_{a_3a_4}(\delta_{\ta_1\ta_3}\delta_{\ta_2\ta_4}
  -\delta_{\ta_1\ta_4}\delta_{\ta_2\ta_3})
  +(\delta_{a_1a_3}\delta_{a_2a_4}-
  \delta_{a_1a_4}\delta_{a_2a_3})\delta_{\ta_1\ta_2}\delta_{\ta_3\ta_4} \\  
\end{split}
\ee
The RHS has the correct antisymmetry under $A_1\leftrightarrow A_2$ or $A_3\leftrightarrow A_4$ as expected from the LHS.
The RHS also satisfies the Jacobi identity. Thus the coset relation is well-supported.

\subsection{G$_{2,1}\oplus$ F$_{4,1}$}

For this case, we choose the factor $(\mathbf{7}, \mathbf{26})$ in the decomposition of the $\mathbf{248}$ of E$_8$ in \eref{decomp}. We start by noting that the algebras $G_2$ and $F_4$ are not simply laced, and thus even at level 1 the blocks will not reduce to elementary functions of $z$. Also, the fusion rules of these theories imply that, for the first time in the present work, there are two distinct blocks contributing to the correlator. 

For G$_{2,1}$, we have $c=\sfrac{14}{5}$ and the single real primary has $h=\frac25$. The correlator of interest is $\langle \mathbf{7}\mathbf{7}\mathbf{7}\mathbf{7}\rangle(z)$. We have the tensor product:
\be
\mathbf{7}\otimes \mathbf{7}=\mathbf{1}\oplus \mathbf{7}\oplus \mathbf{14}\oplus \mathbf{27}
\ee
The $\mathbf{27}$ and $\mathbf{1}$ lie in the symmetric part of the product while the $\mathbf{7}$ and $\mathbf{14}$ lie in the antisymmetric part. Of these, the $\mathbf{14}$
and $\mathbf{27}$ decouple from the theory, but we still have two conformal blocks corresponding to propagation of the conformal family of the $\mathbf{1}$ or the $\mathbf{7}$ in the intermediate channel. For each block there are several sub-blocks corresponding to the actual representation that flows given the tensor structure. Since there are altogether four representations produced by $\mathbf{7}\otimes \mathbf{7}$ we expect that the parameter $N = 3$. Then there are four sub-blocks labelled by $p=0,1,2,3$. 
 
The conformal sub-blocks for $G_{2,1}$ are \cite{Mukhi:2017ugw}:
\begin{equation}
  \begin{split}
    \cF_1^{(p)} (z) &= (z(1-z))^{-\frac45} \,_2F_1\left(p-\sfrac65, \sfrac75-p;\sfrac35;z\right)\\[2mm]
    \cF_2^{(p)} (z) &= \mc N^{(p)} (z(1-z))^{-\frac45}z^{\frac25} \, _2F_1\left(p-\sfrac45, \sfrac95-p;\sfrac75;z\right)\\[3mm]
   | \mc N^{(p)}| &= \frac{\Gamma(-\frac25)}{\Gamma(\frac25)} \sqrt{\frac{\Gamma(\frac{11}{5}-p)\Gamma(p-\frac25)}{\Gamma(\frac75-p)\Gamma(p-\frac65)}}
  \end{split}
  \label{G2.blocks}
\end{equation}
Note that the normalisation factor $\mc N^{(p)}$ is only determined up to a phase. We will comment on this phase later.
The associated tensor structures label the representation that flows in the channel. We first write down the ordering of the tensor structures. We work out the representation flowing in each of the sub-blocks following the procedure explained above and in more detail in \cite{Mukhi:2017ugw}, to find:
\be
\arraycolsep=10pt
\begin{array}{rrrrrrrr}
\cF_1\!:  & D_1^{(0)}\!: &\mathbf{7} &(A) & \qquad \cF_2\!: 
& D_2^{(3)}\!: &\mathbf{1} &(S)\\[2mm]
& D_1^{(1)}\!: &\mathbf{27}	 &(S) && D_2^{(2)}\!: &\mathbf{14}	 &(A) \\[2mm]
& D_1^{(2)}\!: &\mathbf{14}	&(A) && D_2^{(1)}\!: &\mathbf{27} &(S)\\[2mm]
& D_1^{(3)}\!: &\mathbf{1} &(S) && D_2^{(0)}\!: &\mathbf{7} &(A)
\end{array}
\ee
We see from here that $D_1^{(p)}=D_2^{(p)}$. This allows us to drop the subscript and simply write $D^{(p)}$. 

From the symmetry/antisymmetry properties above, we have the relations:
\be
\begin{split}
(D^{(0)}+D^{(2)})_{a_1a_2a_3a_4}
 &= \shalf(\delta_{a_1a_2}\delta_{a_3a_4}-\delta_{a_1a_3}\delta_{a_2a_4})\\
(D^{(1)}+D^{(3)})_{a_1a_2a_3a_4} &= \shalf(\delta_{a_1a_2}\delta_{a_3a_4}+\delta_{a_1a_3}\delta_{a_2a_4})
\end{split}
\label{G2Dsums}
\ee
for the sub-blocks of the first block.

To find the individual $D^{(p)}$ we must use additional information, namely the tensor that fuses $\mathbf{7}\otimes \mathbf{7}$ into the identity in the $(1,4)$ channel, which is just $\delta_{a_1a_4}$, and the tensor  that fuses $\mathbf{7}\otimes \mathbf{7}$ into the $\mathbf{7}$ which we denote $q_{a_1a_4b}$.  With these, we get:
\be
\begin{split}
D^{(3)}&~\sim~ \delta_{a_1a_4}\delta_{a_2a_3}\\
D^{(0)} &~\sim~ q_{a_1a_4b} \,q_{a_2a_3b}
\end{split}
\ee
Thus we have determined all the required invariant tensors. The full conformal blocks are now:
\be
\cF_\alpha^{{\rm G}_{2,1}} = \sum_{p=0}^3 \cF_\alpha^{(p)}(z)D^{(p)}
\ee
We will not explicitly write down the answer, which follows from \eref{G2.blocks}, because it is not particularly illuminating. It is a linear combination of hypergeometric functions and will only simplify after we combine these blocks with the corresponding ones for F$_{4,1}$, to which we now turn.

For F$_{4,1}$ the sole nontrivial primary is the $\mathbf{26}$. The theory has $c=\frac{26}{5}$ and the primary has dimension $\frac35$. The correlator of interest is $\langle \mathbf{26}\, \mathbf{26}\, \mathbf{26}\, \mathbf{26}\rangle(z)$. The relevant product of representations is:
\be
\mathbf{26}\otimes \mathbf{26}=\mathbf{1}\oplus \mathbf{26}\oplus \mathbf{52}\oplus \mathbf{273}\oplus \mathbf{324}
\ee
From the Dynkin labels $(0,0,0,1)$ for the $\mathbf{26}$ and $(0,0,0,2)$ for the
$\mathbf{324}$, we see that the latter is in the symmetric part of the product. Together with the $\mathbf{26}$ and $\mathbf{1}$, this makes up the symmetric part of the product while the $\mathbf{273}$ and $\mathbf{52}$ together make up the antisymmetric part.

In the CFT, all except the first two representations decouple. Thus there are two conformal blocks corresponding to the family of the $\mathbf{1}$ and the $\mathbf{26}$.
Since the representation theory gives rise to five output representations, we expect that $N=4$. Thus there will be five sub-blocks for each conformal block. These are \cite{Mukhi:2017ugw}:
\begin{equation}
  \begin{split}
    \tcF_1^{(p)} (z) &= (z(1-z)^{-\frac65} \,_2F_1\left(p-\sfrac95, \sfrac85-p;\sfrac25;z\right)\\[2mm]
    \tcF_2^{(p)} (z) &= \mc N^{(p)} (z(1-z)^{-\frac65}z^{\frac35} \, _2F_1\left(p-\sfrac65, \sfrac{11}{5}-p;\sfrac85;z\right)\\[2mm]
   | \wtd{\mc N}^{(p)}| &= \frac{\Gamma(-\frac35)}{\Gamma(\frac35)} \sqrt{\frac{\Gamma(\frac{14}{5}-p)\Gamma(p-\frac35)}{\Gamma(\frac85-p)\Gamma(p-\frac95)}}
  \end{split}
\end{equation}
Again the normalisation factor is determined only up to a phase.

Let us now work out the corresponding invariant tensors. The representations flowing in the sub-blocks are found to be:
\be
\arraycolsep=10pt
\begin{array}{rrrrrrrr}
\cF_1\!:  & \tD_1^{(0)}\!: &\mathbf{26} &(S) & \qquad \cF_2\!: & \tD_2^{(4)}\!: &\mathbf{1} &(S)\\[2mm]
& \tD_1^{(1)}\!: &\mathbf{273}	 &(A) && \tD_2^{(3)}\!: &\mathbf{52}	 &(A) \\[2mm]
& \tD_1^{(2)}\!: &\mathbf{324}	&(S) && \tD_2^{(2)}\!: &\mathbf{324} &(S)\\[2mm]
& \tD_1^{(3)}\!: &\mathbf{52} &(A) &&\tD_2^{(1)}\!: &\mathbf{273} &(A)\\[2mm]
& \tD_1^{(4)}\!: &\mathbf{1} &(S) && \tD_2^{(0)}\!: &\mathbf{26} &(S)
\end{array}
\ee
Again $\tD_1^{(p)}=\tD_2^{(p)}$ and we can just write $\tD^{(p)}$. 

It follows that:
\be
\begin{split}
(\tD^{(0)}+\tD^{(2)}+\tD^{(4)})_{\ta_1\ta_2\ta_3\ta_4}
 &= \shalf(\delta_{\ta_1\ta_2}\delta_{\ta_3\ta_4}+\delta_{\ta_1\ta_3}\delta_{\ta_2\ta_4})\\[2mm]
(\tD^{(1)}+\tD^{(3)})_{\ta_1\ta_2\ta_3\ta_4} &= \shalf(\delta_{\ta_1\ta_2}\delta_{\ta_3\ta_4}-\delta_{\ta_1\ta_3}\delta_{\ta_2\ta_4})
\end{split}
\label{F4Dsums1}
\ee

To completely specific the tensor invariants we note that $\delta_{\ta_1\ta_4}$ combines 
$\mathbf{26}\otimes \mathbf{26}$ into $\mathbf{1}$ in the $(1,4)$ channel, while the third-rank tensors $\tq_{\ta_1\ta_4\tb}$ and ${\tilde r}_{\ta_1\ta_4\tmu}$ are defined to be the ones that combine $\mathbf{26}\otimes \mathbf{26}$ into the $\mathbf{26}$ and the $\mathbf{52}$ respectively. Here $\tmu\in 1,\cdots,52$. Then we have:
\be
\begin{split}
\tD^{(4)} & \sim \delta_{\ta_1\ta_4}\delta_{\ta_2\ta_3}\\
\tD^{(3)} & \sim \tilde r_{\ta_1\ta_4\tmu}\,\tilde r_{\ta_2\ta_3\tmu} \\
\tD^{(0)} & \sim \tq_{\ta_1\ta_4\tb}\,\tq_{\ta_2\ta_3\tb} 
\end{split}
\ee
up to normalisation. Together with \eref{F4Dsums1} this determines all the $\tD^{(p)}$.

On combining the blocks for G$_{2,1}$ and F$_{4,1}$ as per our conjecture \eqref{cosetconj}, using results from Appendix \ref{App.Bilinear}, and choosing the phases of the normalisations suitably (as we describe below), we find:
\begin{equation}
  \begin{split}
    &\sum_{\alpha} \bigg(\sum_{p,\tp} D^{(p)}\wtd D^{(\tp)} \cF^{(p)}_\alpha(z) \tcF^{(\tp)}_\alpha(z) \bigg) =
    \sum_{p,\tp} D^{(p)}\wtd D^{(\tp)}  \bigg(\sum_{\alpha} \cF^{(p)}_\alpha(z) \tcF^{(\tp)}_\alpha(z)\bigg) \\    
    &=\frac{1}{(z(1-z))^2}\bigg(\vphantom{\frac12}
    D^{(0)} \wtd D^{(0)} \left(-28 z^3+42 z^2-16 z+1\right)\\
    &\quad +D^{(0)} \wtd D^{(1)} \left(6 z^2-6 z+1\right)+D^{(0)} \wtd D^{(2)} (1-2 z)+D^{(0)} \wtd D^{(3)}+D^{(0)} \wtd D^{(4)} (1-2 z) \\
    &\quad +D^{(1)} \wtd D^{(0)} \left(\frac{28 z^2}{3}-\frac{28 z}{3}+1\right) +D^{(1)} \wtd D^{(1)} (1-2 z)+D^{(1)} \wtd D^{(2)} \\
    &\quad +D^{(1)} \wtd D^{(3)} (1-2 z)+D^{(1)} \wtd D^{(4)} \left(\frac{26 z^2}{3}-\frac{26 z}{3}+1\right) \\
    &\quad +D^{(2)} \wtd D^{(0)} (1-2 z)+D^{(2)} \wtd D^{(1)}+D^{(2)} \wtd D^{(2)} (1-2 z)\\
    &\quad +D^{(2)} \wtd D^{(3)} \left(12 z^2-12 z+1\right)+D^{(2)} \wtd D^{(4)} \left(-52 z^3+78 z^2-28 z+1\right)\\
    &\quad +D^{(3)} \wtd D^{(0)}+D^{(3)} \wtd D^{(1)} (1-2 z)+D^{(3)} \wtd D^{(2)} \left(7 z^2-7 z+1\right) \\
    &\quad  +D^{(3)} \wtd D^{(3)} \left(-42 z^3+63 z^2-23 z+1\right)\\
    &\quad +D^{(3)} \wtd D^{(4)} \left(182 z^4-364 z^3+228 z^2-46 z+1\right)
    \vphantom{\frac12}\bigg)
  \end{split}
  \label{g2.f4.prod}
\end{equation}

In the above, we picked the phases for the normalisation factors to simplify the product of hypergeometric functions appropriately, informed by \eqref{npos}. The solution to this requirement turned out to be:
\begin{equation}
 \mc N^{(p)}= a|\mc N^{(p)}|,\quad  \mc{\wtd N}^{(p)}=a|\mc{\wtd N}^{(p)}|\quad \hbox{where } a^2 = \begin{cases}
    1 & p < 2\\
    -1 &p \geq 2
    \end{cases}
\end{equation}

\Comment{
The following matrix lists the product of the normalizations of the second conformal block, with the sign which allowed the hypergeometric functions to be simplified into polynomials. 
\begin{equation}
  \mc N^{(n)} \mc {\wtd N}^{(m)}= \left(
  \begin{array}{ccccc}
    -6 & -2 & 1 & 7 & 14 \\
    -2 & -\sfrac23 & \sfrac13 & \sfrac73 & \sfrac{14}{3} \\
    6 & 2 & -1 & -7 & -14 \\
    12 & 4 & -2 & -14 & -28 \\
  \end{array}
  \right)
\end{equation}
}

After collecting terms, \eref{g2.f4.prod} can be rewritten:
\begin{equation}
  \begin{split}
    &\frac{(D^{(0)}+D^{(1)}+D^{(2)}+D^{(3)})(\wtd D^{(0)}+\wtd D^{(1)}+\wtd D^{(2)}+\wtd D^{(3)}+\wtd D^{(4)})}{z^2} \\
    &+ \frac{(-D^{(0)}+D^{(1)}-D^{(2)}+D^{(3)})(\wtd D^{(0)}-\wtd D^{(1)}+\wtd D^{(2)}-\wtd D^{(3)}+\wtd D^{(4)})}{(1-z)^2}+ 182 D^{(3)} \wtd D^{(4)}\\
    &+ \frac{1}{z}\left( D^{(0)}\left(-14\wtd D^{(0)}-4\wtd D^{(1)}+2\wtd D^{(3)}\right)+ D^{(1)}\left(-\sfrac{22}{3}\wtd D^{(0)}+2\wtd D^{(2)}-\sfrac{20}{3}\wtd D^{(4)}\right)\right.\\
    &\qquad\left.+D^{(2)}\left(2\wtd D^{(1)}-10\wtd D^{(3)} -26\wtd D^{(4)}\right)+ D^{(3)}\left(2\wtd D^{(0)}-5\wtd D^{(2)}-21\wtd D^{(3)}-44\wtd D^{(4)}\right)\right)\\
    &+ \frac{1}{1-z}\left( D^{(0)}\left(14\wtd D^{(0)}-4\wtd D^{(1)}+2\wtd D^{(3)}\right)+ D^{(1)}\left(-\sfrac{22}{3}\wtd D^{(0)}+2\wtd D^{(2)}-\sfrac{20}{3}\wtd D^{(4)}\right)\right.\\
    &\qquad\left.+D^{(2)}\left(2\wtd D^{(1)}-10\wtd D^{(3)} +26\wtd D^{(4)}\right)+ D^{(3)}\left(2\wtd D^{(0)}-5\wtd D^{(2)}+21\wtd D^{(3)}-44\wtd D^{(4)}\right)\right)
  \end{split}
\label{g2.f4.prod.simp}
\end{equation}
Remarkably the dust has settled and we find the predicted form: double and simple poles at $z=0,1$ and a constant term!

For the coset conjecture to hold, we must have the following identifications coming from the double-pole and constant terms:
\begin{equation}
  \begin{split}
    \delta_{A_1 A_2} \delta_{A_3 A_4} &= (D^{(0)}+D^{(1)}+D^{(2)}+D^{(3)})(\wtd D^{(0)}+\wtd D^{(1)}+\wtd D^{(2)}+\wtd D^{(3)}+\wtd D^{(4)})\\
    \delta_{A_1 A_3} \delta_{A_4 A_2} &= (-D^{(0)}+D^{(1)}-D^{(2)}+D^{(3)})(\wtd D^{(0)}-\wtd D^{(1)}+\wtd D^{(2)}-\wtd D^{(3)}+\wtd D^{(4)})\\
    \delta_{A_1 A_4} \delta_{A_2 A_3} &= 182\, D^{(3)} \wtd D^{(4)}
\end{split}
\ee
as well as the following identification coming from the simple pole at $z=0$:
\be
\begin{split}
    f_{A_1 A_2 B}f_{B A_3 A_4}&= D^{(0)}\left(-14\wtd D^{(0)}-4\wtd D^{(1)}+2\wtd D^{(3)}\right)\\
    &\quad + D^{(1)}\left(-\sfrac{22}{3}\wtd D^{(0)}+2\wtd D^{(2)}-\sfrac{20}{3}\wtd D^{(4)}\right)\\
    &\quad+D^{(2)}\left(2\wtd D^{(1)}-10\wtd D^{(3)} -26\wtd D^{(4)}\right)\\
    &\quad + D^{(3)}\left(2\wtd D^{(0)}-5\wtd D^{(2)}-21\wtd D^{(3)}-44\wtd D^{(4)}\right)
\end{split}
\ee 
Substituting the tensor invariants for G$_2$:
\begin{equation}
  \begin{split}
    D^{(0)} &= N^{(0)} q_{a_1 a_4 b} q_{a_2 a_3 b} \\
    D^{(1)} &= \sfrac12(\delta_{a_1 a_2}\delta_{a_3 a_4}+\delta_{a_1 a_3}\delta_{a_4 a_2}) - N^{(3)} \delta_{a_1 a_4}\delta_{a_2 a_3}\\
    D^{(2)} &= \sfrac12(\delta_{a_1 a_2}\delta_{a_3 a_4}-\delta_{a_1 a_3}\delta_{a_4 a_2}) - N^{(0)} q_{a_1 a_4 b} q_{a_2 a_3 b} \\
    D^{(3)} &= N^{(3)} \delta_{a_1 a_4}\delta_{a_2 a_3}
  \end{split}
\end{equation}
and tensor invariants for F$_4$: 
\begin{equation}
  \begin{split}
    \wtd D^{(0)} &= \wtd N^{(0)} \wtd q_{\ta_1 \ta_4 \tb} \wtd q_{\ta_2 \ta_3 \tb} \\
    \wtd D^{(1)} &= \sfrac12(\delta_{\ta_1 \ta_2}\delta_{\ta_3 \ta_4}-\delta_{\ta_1 \ta_3}\delta_{\ta_4 \ta_2}) - \wtd N^{(3)} \wtd r_{\ta_1 \ta_4\mu}\wtd r_{\ta_2 \ta_3\mu}\\
    \wtd D^{(2)} &= \sfrac12(\delta_{\ta_1 \ta_2}\delta_{\ta_3 \ta_4}-\delta_{\ta_1 \ta_3}\delta_{\ta_4 \ta_2}) - \wtd N^{(0)} q_{\ta_1 \ta_4 b} q_{\ta_2 \ta_3 b} - \wtd N^{(4)} \delta_{\ta_1 \ta_4}\delta_{\ta_2 \ta_3} \\
    \wtd D^{(3)} &= \wtd N^{(3)} \wtd r_{\ta_1 \ta_4 \mu} \wtd r_{\ta_2 \ta_3 \mu} \\
    \wtd D^{(4)} &= \wtd N^{(4)} \delta_{\ta_1 \ta_4}\delta_{\ta_2 \ta_3} \\
  \end{split}
\end{equation}
we get:
\begin{equation}
  \begin{split}
    \delta_{A_1 A_2} \delta_{A_3 A_4} &= \delta_{a_1 a_2} \delta_{a_3 a_4} \delta_{\ta_1 \ta_2} \delta_{\ta_3 \ta_4} \\
    \delta_{A_1 A_3} \delta_{A_4 A_2} &= \delta_{a_1 a_3} \delta_{a_4 a_2} \delta_{\ta_1 \ta_3} \delta_{\ta_4 \ta_2} \\
    \delta_{A_1 A_4} \delta_{A_2 A_3} &= 182\, N^{(3)} \wtd N^{(4)}\delta_{a_1 a_4} \delta_{a_2 a_3} \delta_{\ta_1 \ta_4} \delta_{\ta_2 \ta_3} \\
  \end{split}
\end{equation}
Thus, choosing the product of the relevant normalisations to be:
\be
N^{(3)} \wtd N^{(4)}=\frac{1}{182}
\ee
we find perfect agreement with expectations. 

From the single pole terms,  we find the identifications:
\begin{equation}
  \begin{split}
    &f_{A_1 A_2 B}f_{B A_3 A_4}\\
    &= N^{(0)} q_{a_1 a_4 b} q_{a_2 a_3 b}\left(-14\wtd N^{(0)} \wtd q_{\ta_1 \ta_4 \tb} \wtd q_{\ta_2 \ta_3 \tb} -3(\delta_{\ta_1 \ta_2}\delta_{\ta_3 \ta_4}-\delta_{\ta_1 \ta_3}\delta_{\ta_4 \ta_2})\right.\\
    &\qquad\qquad\left.+ 18\wtd N^{(3)} \wtd r_{\ta_1 \ta_4\mu}\wtd r_{\ta_2 \ta_3\mu} +26\wtd N^{(4)} \delta_{\ta_1 \ta_4}\delta_{\ta_2 \ta_3}\right)\\
    &\quad +\delta_{a_1 a_2}\delta_{a_3 a_4}
    \left(-\sfrac{14}{3}\wtd N^{(0)} \wtd q_{\ta_1 \ta_4 \tb} \wtd q_{\ta_2 \ta_3 \tb} +\delta_{\ta_1 \ta_2}\delta_{\ta_3 \ta_4} -6\wtd N^{(3)} \wtd r_{\ta_1 \ta_4 \mu} \wtd r_{\ta_2 \ta_3 \mu} -\sfrac{52}{3}\wtd N^{(4)} \delta_{\ta_1 \ta_4}\delta_{\ta_2 \ta_3}\right)\\
    &\quad+\delta_{a_1 a_3}\delta_{a_4 a_2}
    \left(-\sfrac{14}{3}\wtd N^{(0)} \wtd q_{\ta_1 \ta_4 \tb} \wtd q_{\ta_2 \ta_3 \tb} + \delta_{\ta_1 \ta_3}\delta_{\ta_4 \ta_2}  +6\wtd N^{(3)} \wtd r_{\ta_1 \ta_4 \mu} \wtd r_{\ta_2 \ta_3 \mu} +\sfrac{26}{3}\wtd N^{(4)} \delta_{\ta_1 \ta_4}\delta_{\ta_2 \ta_3}\right)\\
    &+ N^{(3)} \delta_{a_1 a_4}\delta_{a_2 a_3}\left(
    \sfrac{49}{3}\wtd N^{(0)} q_{\ta_1 \ta_4 b} q_{\ta_2 \ta_3 b} -\sfrac72(\delta_{\ta_1 \ta_2}\delta_{\ta_3 \ta_4}+\delta_{\ta_1 \ta_3}\delta_{\ta_4 \ta_2})\right.\\
    &\qquad\qquad\left. -21\wtd N^{(3)} \wtd r_{\ta_1 \ta_4 \mu} \wtd r_{\ta_2 \ta_3 \mu} -\sfrac{91}{3}\wtd N^{(4)} \delta_{\ta_1 \ta_4}\delta_{\ta_2 \ta_3}\right) \\
  \end{split}\label{g2f4pole0}
\end{equation}
This is unfortunately rather complicated and we will leave its detailed investigation for future work. Nonetheless,  this non-simply-laced case has passed a number of checks that give strong evidence for the coset conjecture. 

\subsection{Intermediate VOA's: A$_{0.5}\oplus$ E$_{7.5}$}

In this subsection we note that similar considerations to those discussed in previous sections appear to apply to Intermediate Vertex Operator Algebras \cite{Kawasetsu:2014}, of which the first two examples form part of the MMS series discovered nearly three decades earlier \cite{Mathur:1988na}. To our knowledge, correlation functions of IVOA's have not been studied in detail. However their fusion rules are known (if one is willing to ignore some negative signs in the fusion coefficients \cite{Mathur:1988gt}) and so one can simply apply the methods of \cite{Mukhi:2017ugw} to them and see what one finds.

The first IVOA is A$_{0.5}$, for which $ c = \sfrac25,\, h = \sfrac15$. This is a presentation of the familiar Lee-Yang minimal model but with the role of the identity and non-trivial primary interchanged. Both fields have no degeneracy. Applying the rules derived in \cite{Mukhi:2017ugw}, we find that there should be three sub-blocks. Given the absence of degeneracy (and hence independent tensor structures) this is a little puzzling. Still, if we apply the universal formula of the above paper to this case, we find the following conformal sub-blocks for the nontrivial 4 point correlator, where $n=0,1,2$:
\begin{equation}
  \begin{split}
    \mc F^{(n)}_1(z) &= (z(1 - z))^{-\sfrac25} \,_2F_1(\sfrac65 - n, -\sfrac35 + n, \sfrac45, z) \\
    \mc F^{(n)}_2(z) &= \mc N^{(n)} (z(1 - z))^{-\sfrac25} z^{\sfrac15}\,_2F_1(\sfrac75 - n, -\sfrac25 + n, \sfrac65, z) \\
    \mc N^{(n)} &= \frac{\Gamma(-\sfrac15)}{\Gamma(\sfrac15)} \sqrt{\frac{\Gamma(\sfrac85 - n) \Gamma(-\sfrac15 + n)}{\Gamma(\sfrac65 - n) \Gamma(-\sfrac35 + n)}}
  \end{split}
\end{equation}

The second theory of the pair is the E$_{7.5,1}$ IVOA. This has $c = \sfrac{38}{5}$ and $h=\sfrac45$. Here the universal formulae give us a set of sub-blocks where $n=0,1,\cdots,6$:
\begin{equation}
  \begin{split}
    \wtd{\mc F}^{(n)}_1(z) &= (z(1 - z))^{-\sfrac85} \,_2F_1(\sfrac95 - n, -\sfrac{12}{5} + n, \sfrac15, z) \\
    \wtd{\mc F}^{(n)}_2(z) &= \wtd{\mc N}^{(n)} (z(1 - z))^{-\sfrac85} z^{\sfrac45}\,_2F_1(\sfrac{13}{5} - n, -\sfrac85 + n, \sfrac95, z) \\
    \wtd{\mc N}^{(n)} &= \frac{\Gamma(-\sfrac45)}{\Gamma(\sfrac45)} \sqrt{\frac{\Gamma(\sfrac{17}{5} - n) \Gamma(-\sfrac45 + n)}{\Gamma(\sfrac95 - n) \Gamma(-\sfrac{12}{5} + n)}}
  \end{split}
\end{equation}

We now test whether a bilinear relation among correlators is possible. For this, we assume there exist some generalised ``tensor structures'' $D^{(p)},\tD^{(p)}$. The first of these should be pure numbers, in the absence of degeneracy, while the second could be tensors related to the Intermediate Lie Algebra E$_{7.5}$ which in turn bears a relationship with E$_7$. Without going into the details of what these numbers/tensors are, we can write out the full conformal blocks and then combine them pairwise, and see whether this can potentially match the current correlators of E$_{8,1}$.

It turns out that this product, after summing over all blocks, does indeed become an elementary function of $z$ -- even though each factor is hypergeometric, as a consequence of \eqref{npos}. For this, the sign of some of the normalisations has to be taken negative (as we saw in the previous section these signs, or more generally phases, are not determined at the outset). Then one finds:
\begin{equation}
  \begin{split}
&\sum_{p,\wtd p} D^{(p)} \wtd D^{(p)} \Big(\sum_{\alpha}\mc F^{(p)}_\alpha \wtd{\mc F}^{(\wtd p)}_\alpha\Big)  \\  
    &=\quad \frac{(D^{(0)} + D^{(1)} + D^{(2)})(\wtd D^{(0)} + \wtd D^{(1)} +\wtd  D^{(2)} + \wtd  D^{(3)} + \wtd D^{(4)} +\wtd  D^{(5)})}{z^2}\\
    &\qquad +\frac{(-D^{(0)} + D^{(1)} - D^{(2)})(\wtd D^{(0)} - \wtd D^{(1)} +\wtd  D^{(2)} - \wtd  D^{(3)} + \wtd D^{(4)} -\wtd  D^{(5)})}{(1-z)^2}+ 57 D^{(2)}\wtd D^{(5)}\\
    &\qquad +\frac1z\Big(
    D^{(0)}(11 \wtd D^{(0)} + 6 \wtd D^{(1)} + 2 \wtd D^{(3)} -\sfrac{15}{2} \wtd D^{(5)})\\
    &\qquad\qquad 
    +D^{(1)}(-9 \wtd D^{(0)} + 2 \wtd D^{(2)} - 10 \wtd D^{(4)} -\sfrac{57}{2} \wtd D^{(5)})\\
    &\qquad\qquad 
    +D^{(2)}(2 \wtd D^{(1)} - 6 \wtd D^{(4)} -19 \wtd D^{(5)})
    \Big)\\
    &\qquad +\frac{1}{1-z}\Big(
    D^{(0)}(-11 \wtd D^{(0)} + 6 \wtd D^{(1)} + 2 \wtd D^{(3)} -\sfrac{15}{2} \wtd D^{(5)})\\
    &\qquad\qquad
    +D^{(1)}(-9 \wtd D^{(0)} +2 \wtd D^{(2)} - 10 \wtd D^{(4)} +\sfrac{57}{2} \wtd D^{(5)})\\
    &\qquad\qquad 
    +D^{(2)}(2 \wtd D^{(1)} + 6 \wtd D^{(4)} -19 \wtd D^{(5)})
    \Big)
  \end{split}
\end{equation}
We see that, remarkably, the result has double and single poles at $z=0,1$ and a constant term. This is highly suggestive that -- despite not being conventional RCFT -- these two IVOA's also fit into the novel coset scheme. We may note that \cite{Hampapura:2016mmz} found evidence of novel coset relations where neither member of the pair was a WZW theory, and the present result seems to confirm that more general cosets should exist. This pair of theories deserves to be examined in more detail in this context. 

\subsection{A$_{4,1}\oplus$ A$_{4,1}$}

This case involves a pair of 3-character theories that were not previously identified as being a coset pair with respect to E$_8$. The primaries of A$_{4,1}$ other than the identity are the $\mathbf{5}$ and $\mathbf{10}$ and their complex conjugates.  
Because these representations are complex, the corresponding characters occur with multiplicity 2 and the partition function is:
\be
Z(\tau,\btau)=|\chi_{\mathbf{1}}(\tau)|^2+2|\chi_{\mathbf{5}}(\tau)|^2+2|\chi_{\mathbf{10}}(\tau)|^2
\ee
where $j^{1/3}$ is the E$_8$ character. 

The central charge and conformal dimensions of the theory are:
\begin{equation}
  c = 4;\quad h_{\mathbf{5}} = \sfrac25;\quad h_{\mathbf{10}} = \sfrac35
\end{equation}
To see the coset relation, we note first that the $\mathbf{5}$ and $\mathbf{10}$ have dimensions that add up to 1. Thus, the bilinear relation requires us to pair the characters of the $\mathbf{5}$ and $\mathbf{\ovl{5}}$ with the characters of the $\mathbf{10}$ and $\mathbf{\ovl{10}}$. This leads to four copies of the product $\chi_{\mathbf{5}}\,\chi_{\mathbf{10}}$. Hence the holomorphic bilinear relation should be:
\begin{equation}
 (\chi_{\mathbf{1}}(\tau))^2 + 4\,\chi_{\mathbf{5}}(\tau)\,\chi_{\mathbf{10}}(\tau)=  j^{1/3}
\end{equation}
We have verified that this relation indeed holds. 

Now we would like to see if there is a similar bilinear relation between conformal blocks. 
From the above considerations (as well as the branching rules in \eref{decomp}), we see that the desired relation is between two sets of six correlators each:
\begin{equation}
\begin{split}
&  \langle\mathbf{5}\,\mathbf{\ovl{5}}\,\,\mathbf{\ovl{5}}\,\mathbf{5}\rangle,~
\langle\mathbf{\ovl{5}}\,\mathbf{5}\,\,\mathbf{5}\,\mathbf{\ovl{5}}\rangle,~
\langle\mathbf{5}\,\mathbf{5}\,\,\mathbf{\ovl{5}}\,\mathbf{\ovl{5}}\rangle,~
\langle\mathbf{\ovl{5}}\,\mathbf{\ovl{5}}\,\,\mathbf{5}\,\mathbf{5}\rangle,~
\langle\mathbf{5}\,\mathbf{\ovl{5}}\,\,\mathbf{5}\,\mathbf{\ovl{5}}\rangle,~
\langle\mathbf{\ovl{5}}\,\mathbf{5}\,\,\mathbf{\ovl{5}}\,\mathbf{5}\rangle,~
\\
&  \langle\mathbf{10}\,\mathbf{\ovl{10}}\,\,\mathbf{\ovl{10}}\,\mathbf{10}\rangle,~
\langle\mathbf{\ovl{10}}\,\mathbf{10}\,\,\mathbf{10}\,\mathbf{\ovl{10}}\rangle,~
\langle\mathbf{10}\,\mathbf{10}\,\,\mathbf{\ovl{10}}\,\mathbf{\ovl{10}}\rangle,~
\langle\mathbf{\ovl{10}}\,\mathbf{\ovl{10}}\,\,\mathbf{10}\,\mathbf{10}\rangle,\\
&
\langle\mathbf{10}\,\mathbf{\ovl{10}}\,\,\mathbf{10}\,\mathbf{\ovl{10}}\rangle,~
\langle\mathbf{\ovl{10}}\,\mathbf{10}\,\,\mathbf{\ovl{10}}\,\mathbf{10}\rangle
\end{split}
\label{5and10}
\end{equation}
So far this is similar to the previous complex example of A$_{2,1}\oplus$ E$_{6,1}$. However, \eref{decomp} says we must pair the $\mathbf{5}$ with {\em both} the $\mathbf{10}$ and $\mathbf{\ovl{10}}$. This produces an enormous profusion of terms. As seen in previous examples, the entire set will be needed to potentially reproduce single-pole terms in the current correlator. However, each combination produces the same double-pole term (or does not have a double pole). So we will restrict ourselves to one of the many possible combinations, and examine only double poles.

For the $\langle\mathbf{5}\,\mathbf{\ovl{5}}\,\,\mathbf{\ovl{5}}\,\mathbf{5}\rangle$, the relevant product relation is:
\be
\mathbf{5}\otimes \mathbf{5}=\mathbf{10}\oplus \mathbf{15} 
\ee
The primary in the latter representation is absent at level 1 and hence the corresponding conformal block decouples. The first block has two sub-blocks:
\begin{equation}
  \begin{split}
    \cF_1^{\mathbf{5}\,(0)} (z) &= \frac{1}{(z(1-z))^{\frac45}} = (z(1-z))^{\frac15} \left(\frac1z + \frac{1}{1-z} \right)\\
    \cF_1^{\mathbf{5}\,(1)} (z) &= \frac{1-2z}{(z(1-z))^{\frac45}} = (z(1-z))^{\frac15} \left(\frac1z - \frac{1}{1-z} \right)
  \end{split}
\end{equation}
corresponding to the tensor structures in this correlator. The antisymmetric combination of $\mathbf{5}$ with itself is the $\mathbf{10}$ which means the primary itself flows. This corresponds to $D^{(0)}$. Meanwhile the symmetric combination is the $\mathbf{15}$, which -- as expected -- appears as a first-level secondary above the $\mathbf{10}$, and corresponds to $D^{(1)}$. Thus we have:
\be
\begin{split}
D^{(0)}\!\!: &\qquad \mathbf{\ovl{15}}~~~~~{\rm (S)}\\  
D^{(1)}\!\!: &\qquad \mathbf{\ovl{10}}~~~~~{\rm (A)}
\end{split}
\ee
Because these are the complete symmetric and antisymmetric parts of the product, we can immediately write them as:
\be
\begin{split}
  D^{(0)}_{a_1 \bar a_2 \bar a_3 a_4} &= \shalf(\delta_{a_1\bar a_2}\delta_{\bar a_3a_4} +\delta_{a_1\bar a_3}\delta_{\bar a_2a_4})\\
  D^{(1)}_{a_1 \bar a_2 \bar a_3 a_4} &= \shalf(\delta_{a_1\bar a_2}\delta_{\bar a_3a_4} -\delta_{a_1\bar a_3}\delta_{\bar a_2a_4})\\
\end{split}
\label{D0D1.SU5}
\ee
The complete conformal block is: 
\begin{equation}
\begin{split}
    \cF_1^{\rm A_{4,1},\mathbf{5}}(z) &= \sum_{p=0}^1 \cF_1^{\mathbf{5}\,(p)}D^{(p)}\\
    &=(z(1-z))^{\frac15}\left( D^{(0)}\left(\frac{1}{z}+\frac{1}{1-z}\right) + D^{(1)}\left(\frac{1}{z}-\frac{1}{1-z}\right) \right)
\end{split}
\end{equation}

To compute $\langle\mathbf{10}\,\mathbf{\ovl{10}}\,\,\mathbf{\ovl{10}}\,\mathbf{10}\rangle$ we note the tensor product decomposition:
\begin{equation}
    \mathbf{10}\otimes \mathbf{10} = \mathbf{\overline 5}\oplus \mathbf{\overline{45}}\oplus\mathbf{\overline{50}}
\end{equation}
Since there are 3 representations in the decomposition we will have three independent tensor structures.

With these values of conformal dimensions, on computing the normalization we see that the second block decouples for this case as well. The surviving block has three sub-blocks:
\begin{equation}
  \begin{split}
    \tcF^{\mathbf{10}\,(0)}_1 &= \frac{1}{(z(1-z))^{\frac65}} = (z(1-z))^{-\frac15} \left(\frac1z + \frac{1}{1-z} \right)\\
    \tcF^{\mathbf{10}\,(1)}_1  &= \frac{1-2z}{(z(1-z))^{\frac65}} = (z(1-z))^{-\frac15} \left(\frac1z - \frac{1}{1-z} \right)\\
   \tcF^{\mathbf{10}\,(2)}_1  &= \frac{1+20z-20z^2}{(z(1-z))^{\frac65}} = (z(1-z))^{-\frac15} \left(\frac1z + \frac{1}{1-z} + 20\right)\\
  \end{split}
\end{equation}
In the $z_1, z_4$ channel, the primary flowing is the symmetric representation $\ovl{\mathbf{5}}$.
The representations at descendant levels 1 and 2 are $\ovl{\mathbf{45}}$ and $\ovl{\mathbf{50}}$, which are antisymmetric and symmetric respectively.
Therefore $\tD^{(2)}$ is associated with $\ovl{\mathbf{5}}$, $\tD^{(1)}$ is associated to $\ovl{\mathbf{45}}$ and $\tD^{(0)}$ is associated to $\ovl{\mathbf{50}}$. Thus we have:
\be
\begin{split}
\tD^{(0)}\!\!: &\qquad \mathbf{\ovl{50}}~~~~~{\rm (S)}\\  
\tD^{(1)}\!\!: &\qquad \mathbf{\ovl{45}}~~~~~{\rm (A)}\\  
\tD^{(2)}\!\!: &\qquad \mathbf{~~\ovl{5}}~~~~~{\rm (S)}
\end{split}
\ee
As we have seen several times previously, the above allows us to conclude that:
\be
\begin{split}
(\tD^{(0)}+\tD^{(2)})_{\ta_1 \bta_2 \bta_3 \ta_4} &= \shalf(\delta_{\ta_1\bta_2}\delta_{\bta_3\ta_4}+\delta_{\ta_1\bta_3}\delta_{\bta_2\ta_4})\\
\tD^{(1)}_{\ta_1 \bta_2 \bta_3 \ta_4} &= \shalf(\delta_{\ta_1\bta_2}\delta_{\bta_3\ta_4}-\delta_{\ta_1\bta_3}\delta_{\bta_2\ta_4})
\end{split}
\label{D0D1D2.SU5}
\ee
Thus it only remains to find $\tD^{(2)}$. Let us write the invariant that couples two $\mathbf{10}$'s to a $\mathbf{\ovl{5}}$ as $\tq_{\ta_a\ta_2 b}$. The fourth rank invariant $\tD^{(2)}$ is then:
\be
\tD^{(2)}_{\ta_1 \bta_2 \bta_3 \ta_4}~\sim~\tq_{\ta_1\ta_4 b}\,\btq_{\bb\bta_2\bta_3}
\ee
and the full conformal block is:
\begin{equation}
  \begin{split}
    \tcF^{{\rm A}_{4,1},\mathbf{10}}(z) &=\sum_{p=0}^2\tcF_1^{(p)}\tD^{(p)}\\
 &    \hspace{-1cm} =(z(1-z))^{-\frac15} \left( \frac{\tD^{(0)} + \tD^{(1)} + \tD^{(2)}}{z} + \frac{\tD^{(0)} - \tD^{(1)}+ \tD^{(2)}}{1-z} + 20 \tD^{(2)} \right) 
  \end{split}
\end{equation}

We can now address our conjecture. Multiplying the blocks we have just computed, we get:
\begin{equation}
  \begin{split}
&
\langle\mathbf{5}\,\mathbf{\ovl{5}}\,\,\mathbf{\ovl{5}}\,\mathbf{5}\rangle
\langle\mathbf{10}\,\mathbf{\ovl{10}}\,\,\mathbf{\ovl{10}}\,\mathbf{10}\rangle =\\
    & \frac{(D^{(0)}+D^{(1)}) (\tD^{(0)}+\tD^{(1)} + \tD^{(2)})}{z^2} + \frac{(D^{(0)}-D^{(1)})(\tD^{(0)} - \tD^{(1)} + \tD^{(2)})}{(1-z)^2} \\
    &\quad + \frac{2(D^{(0)} \tD^{(0)} - D^{(1)} \tD^{(1)} + D^{(0)} \tD^{(2)}) + 20(D^{(0)}+D^{(1)}) \tD^{(2)}}{z} \\
    &\quad + \frac{2(D^{(0)} \tD^{(0)} - D^{(1)} \tD^{(1)} + D^{(0)} \tD^{(2)}) + 20(D^{(0)}-D^{(1)}) \tD^{(2)}}{1-z} 
  \end{split}
\end{equation}

Comparing the double-pole terms to those in the E$_{8,1}$ current correlator, we can make the following identifications:
\begin{equation}
  \begin{split}
    (D^{(0)}+D^{(1)})_{a_1\ba_2\ba_3 a_4}
    &=p\,\delta_{a_1\ba_2}\delta_{\ba_3a_4}, \quad 
    \tD^{(0)}+\tD^{(1)}+\tD^{(2)} = p^{-1}\delta_{\ta_1\bta_2}\delta_{\bta_3\ta_4}\\
    (D^{(0)}-D^{(1)})_{a_1\ba_2\ba_3 a_4} &= q\,\delta_{a_1a_3}\delta_{a_2a_4}, \quad 
    \tD^{(0)}-\tD^{(1)}+\tD^{(2)} = q^{-1}\delta_{\ta_1\ta_3}\delta_{\ta_2\ta_4}
  \end{split}
\end{equation}
From Eqs.(\ref{D0D1.SU5}, \ref{D0D1D2.SU5}) we see that the above equations are satisfied with $p=q=1$. 

We note that there is no double pole at infinity. This is consistent with the fact that the first and last representations ($\mathbf{5}$ and $\mathbf{5}$, as well as $\mathbf{10}$ and $\mathbf{10}$) cannot fuse into the identity.

We see that single poles do arise as desired, but cannot check them in detail without adding all possible terms arising from \eref{5and10}. Hence we will leave this case here, with the observation that (i) it has passed some tests, (ii) this is the unique example studied here that does not belong to the MMS series.

\section{Conclusions and Open Problems}

In this work we have found substantial evidence for a holomorphic bilinear relation between conformal blocks for primary four-point functions of pairs of theories related by a coset relation. This mirrors the bilinear relation between the characters of the coset pairs.

It is intriguing that this proposed relation between conformal blocks of different theories seems to work out in different ways for different pairs. For the simply laced cases at level 1 it is well-known that the blocks are elementary functions. However they have specific fractional powers of $z, 1-z$ which neatly cancel when we combine the blocks with those of their coset partners. Thereafter the result has double and simple poles at $z=0,1$ along with a constant term corresponding to a pole at infinity. This fits well with the structure of current correlators in the E$_{8,1}$ theory and in each case we were able to provide very specific, detailed evidence (though not a complete proof) that the proposed relation is true. For the non-simply-laced case it was a miracle to start with, that their hypergeometric conformal blocks simplify into rational functions when we multiply them pairwise. 

This result puts on a firmer footing the proposal of \cite{Gaberdiel:2016zke} that one can define ``novel'' cosets of meromorphic CFT's. However, it is only a first step. The parent CFT for which we have tested the relation is E$_{8,1}$, which is both meromorphic and a WZW model. Most meromorphic CFT's are not WZW models and it is for them that the coset relation is truly novel. A key motivation for the present work is to set the stage for the investigation of families of 2-character RCFT's whose existence was conjectured in \cite{Naculich:1988xv,Hampapura:2015cea} and proved in \cite{Gaberdiel:2016zke}. These are the unique two-character theories with Wronskian index $\ell=2$, and their correlators have not been studied so far. Moreover they obey bilinear relations with conventional WZW models to pair up to meromorphic CFT's with $c=24$, which were classified in \cite{Schellekens:1992db}. These meromorphic theories have chiral algebras of various spins $\ge 2$. In these coset pairs, the dimensions of the primaries add up to 2, rather than 1 as in the case considered here, so one expects bilinear relations that generate correlators of higher-spin currents. We hope to report on this in the future.

Finally, the classification of fermionic rational CFT via modular linear differential equations has recently been initiated in \cite{Bae:2020xzl}. This work extensively discusses bilinear relations among pairs of theories that relate them to meromorphic superconformal theories. These are fermionic analogues of the novel coset relation among characters. The results we have presented here suggest that similar bilinear relations should hold for conformal blocks of suitable correlators in the fermionic theories, relating them to correlators of holomorphic (super)-currents.

\label{conclusionsec}

\section*{Acknowledgements}

RP would like to thank Raj Patil and Palash Singh for fruitful discussions.
This work was initiated while RP was at IISER Pune and he would like to acknowledge the Institute as well as the INSPIRE Scholarship for Higher Education, Government of India. We are grateful for support from a grant by Precision Wires India Ltd. for String Theory and Quantum Gravity research at IISER Pune.

\section*{Appendices}

\appendix

\section{Bilinear Identities for Hypergeometric Functions} 

\label{App.Bilinear}

In this section we prove a class of identities involving products of hypergeometric functions ${}_2 F_1$. These functions are defined in terms of the Pochhammer symbol:
\be
(a)_k\equiv\frac{\Gamma(a+k)}{\Gamma(a)} 
\ee
as:
\be
{}_2 F_1 \left(\begin{matrix}a,b\\c\end{matrix};z\right)\equiv \sum_{k=0}^\infty\frac{(a)_k\,(b)_k}{(c)_k}\frac{z^k}{k!}
\ee
Then, for each integer $n\ge 0$, we have the following bilinear identity:
\begin{equation}
  \begin{split}
    &{_2}F_1\left(\begin{matrix}a,b\\c\end{matrix};z\right) {_2}F_1\left(\begin{matrix}-a-n, n-b\\1-c\end{matrix};z\right) \\
         &+\frac{(a)_{n+1} (1-n+b-c)_n \,z}{(b-n+1)_{n-1} (1+a-c)_n c(c-1)} \times \\
&\qquad\qquad\qquad     {_2} F_1\left(\begin{matrix}1-c+a, 1-c+b\\2-c\end{matrix};z\right) {_2} F_1\left(\begin{matrix}c-a-n, n-b+c\\1+c\end{matrix};z\right) \\
             &= \text{ Degree } n \text{ polynomial of } z
  \end{split} \label{npos}
\end{equation}
The polynomial is easily evaluated for any particular $n$. 

To prove this relation, we note that $_2 F_1 \left(\begin{matrix}a,b\\c\end{matrix};z\right)$ is meromorphic in the parameter $c$, with simple poles at non-positive integers.
Therefore, the LHS of \eqref{npos} has simple poles for all $c\in \mathbb{Z}$. 
Using the following properties of the Pochhammer symbol:
\begin{equation}
  (a)_n (a+n)_m = (a)_{n+m}\qquad (-a)_n = (-1)^m (a-n+1)_n
\end{equation}
one can show that all the residues at $c \in \mathbb{Z}$ cancel for all $n \geq 0$. 
The second term on the LHS of \eqref{npos} additionally has simple poles at $c=1+a+m$ with $m \in \{0,1,\cdots,n-1\}$ and simple poles at $b=m'$ with $m' \in \{1,2,\cdots, n-1\}$. 
These poles are not cancelled by anything in the first term, and therefore must be present on the RHS. 

The residue of the LHS of \eqref{npos} at $c=1+a+m$ is easily shown to be:
\begin{equation}
  \begin{split}
    &\frac{(a)_{n+1} (b-a-n-m)_n}{(b-n+1)_{n-1} (1+a+m)(a+m)} \prod_{\substack{l=0\\m\neq l}}^{n-1}\frac{1}{l-m}\\
    \times\, z&\sum_{k=0}^{m} \sum_{k'=0}^{n-m-1} \frac{(-m)_{k}(1+m-n)_{k'}}{k!\,k'!} \frac{(b-a-m)_{k} (1+a-b+m+n)_{k'}} {(1+a-m)_{k}(2+a+m)_{k'} } z^{k+k'}
  \end{split}
\end{equation}
which is a polynomial of degree $n$. Similarly, the residue of the LHS at
$b = m'$ is:
\begin{equation}
  \begin{split}
    &\frac{(a)_{n+1} (1-n+m'-c)_n }{ (1+a-c)_n c(c-1)} \prod_{\substack{l=0\\ m'\neq n-l-1}}^{n-2}\frac{1}{m'-n+1+l}\\
    \times\,z&(1-z)\sum_{k=0}^{m'-1} \sum_{k'=0}^{n-m'-1} \frac{(1-m')_{k}(1+m'-n)_{k'}}{k!\,k'!} \frac{(1-a)_{k} (1+a+n)_{k'}} {(2-c)_{k}(1+c)_{k'} } z^{k+k'}
  \end{split}
\end{equation}
which is again a polynomial of degree $n$ again.  
The polynomial can be identified by multiplying the residue with its corresponding pole and adding all of them.
For small $n$ it is easier to just expand the LHS in $z$ and keep terms to order $z^n$. By our proof, the terms of order $z^{n+1}$ and higher all vanish.

As examples, the polynomials for $n=0,1,2$ are:
\be
\begin{split}
& n=0\!:  \quad1\\
&n=1\!: \quad 1 - \frac{a-b+1}{a-c+1}z\\
&n=2\!: \quad 1+\frac{(a-b+2) \left(a^2+a (5-3 b)+2 (b-1) (c-2)\right)}{(b-1) (a-c+1)
(a-c+2)}z\\
&\qquad\qquad\qquad +\frac{\left(-a^3+3 a^2 (b-2)+a \left(-3 b^2+12 b-11\right)+b^3-6 b^2+11 b-6\right)}{(b-1) (a-c+1) (a-c+2)}z^2
\end{split}
\ee

\section{Bilinear Relation for Characters}

\label{App.Bilinear.Char}

The special case of the identity \eref{npos} for $n=0$ is: 
\begin{equation}
\begin{split}
 & {_2}F_1\left(\begin{matrix}a, b\\ c\end{matrix};z\right) {_2} F_1\left(\begin{matrix}-a, -b\\1-c \end{matrix};z\right) + \\
  &\qquad\qquad \frac{abz}{c(c-1)}
    {_2}F_1\left(\begin{matrix}1+a-c, 1+b-c \\ 2-c\end{matrix};z\right){_2}F_1\left(\begin{matrix}c-a, c-b\\ 1+c\end{matrix};z\right) = 1 
\end{split}
\label{n0}
\end{equation}
We now apply this to the characters of $\ell = 0$ two-character CFTs, which were originally computed as hypergeometric functions in \cite{Mathur:1988gt,Naculich:1988xv}. The form that will be most useful to us can be found in Eq.(4.3) of \cite{Gaberdiel:2016zke} (we hope the central charge $c$ in the following formulae will not be confused with the parameter $c$ in the preceding identities):
\begin{equation}
  \begin{split}
    \chi_0(\tau) &= j^{\sfrac{c}{24}}\,{_2}F_1 \left(\begin{matrix} \sfrac{1}{12}-\sfrac{h}{2},\sfrac{5}{12}-\sfrac{h}{2}\\ 1-h\end{matrix};\frac{1728}{j}\right)\\
    \chi_1(\tau) &= \sqrt{m}\,j^{\sfrac{c}{24}-h}\,{_2}F_1 \left(\begin{matrix} \sfrac{1}{12}+\sfrac{h}{2},\sfrac{5}{12}+\sfrac{h}{2}\\ 1+h\end{matrix};\frac{1728}{j}\right)
  \end{split}
\label{twochar.char}
\end{equation}
where:
\begin{equation}
  \sqrt{m} = (1728)^h \sqrt{\frac{\mfs(\sfrac{1}{12}-\sfrac{h}{2}) \mfs(\sfrac{5}{12}-\sfrac{h}{2})}{\mfs(\sfrac{1}{12}+\sfrac{h}{2})\mfs(\sfrac{5}{12}+\sfrac{h}{2})}}
  \frac{\Gamma(1-h) \Gamma(\sfrac{11}{12}+\sfrac{h}{2})\Gamma(\sfrac{7}{12}+\sfrac{h}{2})}{\Gamma(1+h)\Gamma(\sfrac{11}{12}-\sfrac{h}{2})\Gamma(\sfrac{7}{12}-\sfrac{h}{2})}
\end{equation}
Here $j$ is the Klein-$j$ invariant and $\mfs(x)\equiv \sin(\pi x)$. 

Recall that the bilinear relation between coset pairs with respect to E$_{8,1}$, which we wish to prove, is:
\begin{equation}
  \chi_0(\tau) \wtd \chi_0(\tau) + \chi_1(\tau) \wtd \chi_1(\tau) = j(\tau)^{\sfrac13}
\label{bil.coset.app}
\end{equation}
Recall \cite{Gaberdiel:2016zke} that the sum of holomorphic dimensions for the coset pairs we are considering is 1, while the sum of central charges is 8.

To prove the above relation, we insert \eref{twochar.char} into \eref{bil.coset.app}. A common factor $j^\frac13$ can now be cancelled from the equation. Next, the product $\sqrt{m\wtd m}$ can be simplified using properties of $\Gamma$ functions. Finally, we use \eref{n0} after making the following identifications:
\begin{equation}
    a=  \frac{1}{12} - \frac{h}{2}, \quad   b=  \frac{5}{12} - \frac{h}{2}, \quad  c= 1-h,\qquad    z = \frac{1728}{j}
\end{equation}
to get the desired result.

\bibliographystyle{JHEP}
\bibliography{cosetcorr}

\end{document}